\documentclass[aps,prd,eqsecnum,showpacs,amsmath,amssymb,superscriptaddress,nofootinbib,preprint]{revtex4-2}

\usepackage{graphicx}
\usepackage{color}
\usepackage{bm}

\newcommand{\hrf}[1]{\hbox to #1{\hrulefill}} 

\newcommand{\suml}{\sum\limits}

\newcommand{\Lc}{\mathcal{L}} 
\newcommand{\diag}{\mathop{\rm diag}\nolimits}

\begin{document}
\title{Cosmological dynamics in the theory of gravity both with non-minimal and non-minimal derivative coupling}

\author{Ravil R. Fatykhov}
\email{frr1802@yandex.ru}
\affiliation{Institute of Physics, Kazan Federal University, Kremliovskaya street 16a, Kazan 420008, Russia}

\author{Sergey V. Sushkov} 
\email{sergey_sushkov@mail.ru}
\affiliation{Institute of Physics, Kazan Federal University, Kremliovskaya street 16a, Kazan 420008, Russia}

\begin{abstract}
This paper explores cosmological scenarios in a scalar-tensor theory of gravity, including both a non-minimal coupling with scalar curvature of the form $R\phi^2$ and a non-minimal derivative coupling of the form $G^{\mu\nu}\phi_{,\mu}\phi_{,\nu}$ in the presence of a scalar field potential with the monomial dependence $V(\phi) = V_0\phi^n$. Critical points of the system were obtained and analyzed. In the absence of a scalar field potential, stability conditions for these points were determined. Using methods of dynamical systems theory, the asymptotic behavior of the model was analyzed. It was shown that in the case of $V(\phi)\equiv0$ or $n < 2$, a quasi-de Sitter asymptotic behavior exists, corresponding to an early inflationary universe. This asymptotic behavior in the approximation $V_0 \rightarrow 0,\ \xi \rightarrow 0$ coincides with the value $H = \frac{1}{\sqrt{9|\eta|}}$ obtained in works devoted to cosmological models with non-minimal kinetic coupling. For $|\xi|\ \rightarrow \infty$, this asymptotic behavior tends to the value $H = \frac{1}{\sqrt{3|\eta|}}$. Moreover, unstable regimes with phantom expansion $w_{eff} < -1$ were found for the early dynamics of the model. For the late dynamics, the following stable asymptotic regimes were obtained: a power-law expansion with $w_{eff} \ge 1$, an expansion with $w_{eff} =\frac{1}{3}$ ($V(\phi)\equiv0$), at which the effective Planck mass tends to zero, and an exponential expansion with $w_{eff} = 0$ as $n = 2$. In this case, the asymptotic value of the Hubble parameter depends only on $V_0 = \frac{1}{2}m^2$ and $\xi$. Numerical integration of the model dynamics was performed for specific values of the theory parameters. The results are presented as phase portraits.
\end{abstract}


\maketitle

\section*{Introduction}

Almost immediately after its creation, the general theory of relativity (GR) was applied to describe the global evolution of the Universe. Using Einstein's equations, Friedmann predicted the expansion of the Universe, which was later brilliantly confirmed (the Hubble expansion law). Thus, GR became the theoretical foundation of cosmology. However, Friedmann's model fails to explain important observable properties of the Universe: problems of flatness, the horizon, and other issues remain unresolved. To resolve these problems, the Universe must have undergone a colossal expansion in its early stages, called inflation \cite{Guth:1980zm, Linde:1982uu}. Importantly, within the framework of Friedmann's model (where the source of gravity is ordinary matter), the expansion of the Universe follows a power law with deceleration. Thus, it was necessary to develop a mechanism that would lead to exponential expansion in the early stages and with subsequent transition to a conventional power-law expansion. The simplest and most developed implementation of this is the introduction of a scalar field (inflaton), minimally coupled to curvature. However, modern astronomical observations \cite{Planck:2018jri, DES:2018paw} indicate that the Universe has relatively recently entered a new phase of accelerated expansion. The substance driving the observed accelerated expansion is called ``dark energy.'' Thus, the above circumstances (the fact of accelerated expansion of the Universe in the past and present) compel us to seek an explanation for the observed cosmological dynamics.

There are two main approaches to describing the modern accelerated expansion of the universe: introducing into Einstein's equations an unknown form of matter such that the pressure $P$ and density $\rho$ of a perfect fluid filling the universe satisfy the relation $P < -\rho/3$ (while modern observations indicate that $P \approx -\rho$), or modifying the Einstein equations themselves. The simplest mechanism for accelerated expansion at a late stage is obtained by adding a term of the form $\Lambda g_{\mu\nu}$ to the left-hand side of Einstein's equations, where $\Lambda$ is the so-called cosmological constant (see review \cite{Carroll:2000fy}). In this case, the $\Lambda$ term can also be regarded as matter with the equation of state $P = -\rho$.

In modern theoretical cosmology, many theories of gravity have been proposed that could explain the accelerated expansion of the universe in the past and/or present (see review \cite{CANTATA:2021asi}). 
A significant number of them are based on the assumption of an additional gravitational degree of freedom. Theories in which gravity is described by a scalar field in addition to a tensor metric field are called scalar-tensor theories. Beyond theories with a scalar field minimally coupled to gravity, nothing a priori prohibits the consideration of theories with non-minimal coupling. For example, the conclusions of the canonical scalar field theory with a quartic potential $\frac{\lambda}{4}\phi^4$ are not satisfactory (at the $2\sigma$ level) for the constraints obtained by the WMAP mission \cite{Planck:2018jri}, which is an additional argument for seriously considering theories with non-minimal coupling.

Within the framework of scalar-tensor theories, there are two main approaches to modifying Einstein's theory of gravity with a canonical scalar field: modifying the ``scalar'' part and introducing an explicit coupling between the tensor and scalar degrees of freedom. It is desirable that the equations of motion obtained from this theory be no higher than second order to avoid ``ghost'' instabilities.

The most general scalar-tensor theory of gravity, leading to second-order equations of motion, is Horndeski's theory \cite{Horndeski:1974wa}. The action for the gravitational part of Horndeski's theory can be written as follows \cite{Kobayashi:2011nu, Kobayashi:2019hrl}:

\[
S = \int d^4x\sqrt{-g}\suml_{i=2}^5\Lc_i,
\]

\noindent where

\begin{align*}
	\Lc_2 &= G_2(\phi,X), \\
	\Lc_3 &= -G_3(\phi,X)\square\phi, \\
	\Lc_4 &= G_4(\phi,X)R + G_{4X}(\phi,X)\left[ (\square\phi)^2 - (\phi_{;\mu\nu})^2 \right], \\
	\Lc_5 &= G_5(\phi,X)G^{\mu\nu}\phi_{;\mu\nu} - \frac{1}{6}G_{5X}(\phi,X)\left[ (\square\phi)^3 + 2(\phi_{;\mu\nu})^3 - 3(\phi_{;\mu\nu})^2\square\phi \right].
\end{align*}

Here $G_i$ are arbitrary functions of the scalar field $\phi$ and $X$, $G_{iX} \equiv \frac{\partial G_i}{\partial X}$, $\phi_{,\alpha} \equiv \nabla_\alpha\phi$, $\phi_{;\mu\nu} \equiv \nabla_\nu\nabla_\mu\phi$ are covariant derivatives, $X \equiv -\frac{1}{2}(\nabla\phi)^2 = -\frac{1}{2}\phi^{,\alpha}\phi_{,\alpha}$ is the kinetic term of the field $\phi$, $\square\phi \equiv \phi^{;\alpha}{}_\alpha$, $(\phi_{;\mu\nu})^2 \equiv \phi_{;\mu\nu}\phi^{;\mu\nu}$, $(\phi_{;\mu\nu})^3 \equiv \phi_{;\mu}{}^\nu\phi_{;\nu}{}^\alpha\phi_{;\alpha}{}^\mu$, $R$ and $G_{\mu\nu}$ are the Ricci scalar and the Einstein tensor respectively.

It can be seen that the terms $\Lc_2$ and $\Lc_3$ modify the scalar part and are expressed only in terms of the field $\phi$ and its derivatives of the form $(\nabla\phi)^2$ and $\square\phi$, while the terms $\Lc_4$ and $\Lc_5$ include an explicit (non-minimal) coupling to the curvature via the Ricci scalar $R$ and the Einstein tensor $G_{\mu\nu}$, respectively. Moreover, the terms $\Lc_4$ and $\Lc_5$ contain corrections to the scalar part to ensure that the order of the equations of motion remains no higher than second. It is worth noting that the terms $G_4 = X$ and $G_5 = -\phi$, as well as $G_3 = f(\phi)$ and $G_2 = -2Xf_\phi$, coincide up to a total derivative and are thus equivalent in terms of the resulting equations of motion \cite{Kobayashi:2011nu}.

In the special case of $\Lc_4$ theories of the form $f(\phi)R$, they are usually called non-minimally coupled theories. This subclass of scalar-tensor theories has been intensively studied in the literature (see, for example, reviews \cite{Fujii2003, Faraoni2004}). It includes, among other theories, the first scalar-tensor theory of gravity proposed by Brans and Dicke \cite{Brans1961}. After renaming, this theory can be reduced to a form in which $f(\phi) = \phi^2$, and the field $\phi$ plays the role of the effective Planck mass. Inflationary cosmology has also developed theories in which both terms are present: the Hilbert--Einstein term $M^2_{\text{Pl}}R$ and the term describing the non-minimal coupling of the scalar field with the curvature $\xi R\phi^2$. With this choice of the non-minimal coupling, the parameter $\xi$ turns out to be dimensionless. Including the term $\xi\phi^2R$ in the action along with the canonical action for the scalar field can lead to an inflationary stage satisfying observational constraints if $\xi\phi^2 > M^2_{\text{Pl}}$ \cite{Komatsu1999}. 
However, this is an undesirable property from the point of view of quantum theory, since it leads to a violation of unitarity \cite{Bezrukov2008}. 

Another special case of non-minimal coupling to curvature are theories with $\Lc_5$ of the form $\eta\phi G_{\mu\nu}\phi^{;\mu\nu}$, which are usually called theories with non-minimal derivative coupling \cite{Amendola1993}. By integration by parts, this term can be reduced to $\eta G^{\mu\nu}\phi_{,\mu}\phi_{,\nu}$, where the parameter $\eta$ has dimensions $M^{-2}$. The presence of this term in the theory leads to interesting features for cosmology. The non-minimal derivative coupling term exerts an additional damping effect on the scalar field dynamics beyond the standard Hubble friction, and, depending on the sign of $\eta$, can lead either to an inflationary stage without introducing a scalar field potential \cite{Sushkov2009, Saridakis2010, Sushkov2012}, or to a slow-rolling stage when considering a scalar field potential. It was shown \cite{Amendola1993} that theories with derivative coupling cannot be transformed by a conformal transformation of the metric to an Einstein theory with a minimally coupled scalar field.

Further interest in theories with non-minimal coupling of the form $\xi R\phi^2$ was stimulated by the papers \cite{Bezrukov2008, Bezrukov2011}, which showed that the Standard Model Higgs field can successfully act as an inflaton field, as well as by the paper \cite{Germani2010}, which considered Higgs inflation in a theory with non-minimal derivative coupling of the form $G^{\mu\nu}\phi_{,\mu}\phi_{,\nu}$.

The paper \cite{Avdeev2022} analyzed non-minimal derivative coupling in the presence of a power-law potential. The paper \cite{Dalianis2020} investigated an inflationary model with non-minimal derivative coupling of the generalized form $G(\phi)G^{\mu\nu}\phi_{,\mu}\phi_{,\nu}$. For numerical integration, the $G(\phi)$ function was chosen in monomial form, and the potential was chosen in monomial and exponential forms. In \cite{Karydas2021}, cosmological scenarios were studied in the presence of both types of non-minimal couplings of the generalized form $R + f(\phi)R$ and $G(\phi)G^{\mu\nu}\phi_{,\mu}\phi_{,\nu}$ in the context of Higgs inflation. For numerical integration, the $f(\phi)$ function was chosen in the standard form $\xi\phi^2$, and the $G(\phi)$ function was also chosen as a monomial. In these studies, stability analysis with respect to metric perturbations was performed.

Cosmological scenarios with non-minimal and non-minimal derivative couplings have been studied from the perspective of dynamical systems theory in a number of papers. The model $F(\phi) = 1 - \xi\phi^2$ was studied in the paper \cite{Szydlowski2009, Hrycyna2010, Szydlowski2014}, where in the case of a general potential, solutions were described characterized by. In the papers of Granda \cite{Granda2012c, Granda2016}, cosmological dynamics was studied in the theory of gravity with a non-minimal kinetic term $G^{\mu\nu}\phi_{,\mu}\phi_{,\nu}$ and a Gauss-Bonnet term $\mathcal{G} = R^2 - 4R_{\mu\nu}R^{\mu\nu} + R_{\mu\nu\rho\sigma}R^{\mu\nu\rho\sigma}$ with an exponential potential $V(\phi) = V_0e^{\kappa\phi}$. Cosmological dynamics with power-law potential $V(\phi) = V_0\phi^n$ was analyzed in the papers \cite{Sami2012, Skugoreva2014} for the model $F(\phi) = 1 - \xi\phi^N$ and in the paper \cite{Skugoreva2013} in the presence of a non-minimal derivative coupling $G^{\mu\nu}\phi_{,\mu}\phi_{,\nu}$. In the papers \cite{Matsumoto2015, Matsumoto2018}, cosmological dynamics with non-minimal derivative coupling was studied in the presence of a Higgs-type potential with broken symmetry $V(\phi) = \frac{\lambda}{4}(\phi^2-v^2)^2$ and an exponential potential.


In this paper, we will explore cosmological scenarios in a scalar-tensor theory of gravity, which includes both a non-minimal coupling with scalar curvature of the form $R\phi^2$ and a non-minimal derivative coupling of the form $G^{\mu\nu}\phi_{,\mu}\phi_{,\nu}$ in the presence of a scalar field potential with the monomial dependence $V(\phi) = V_0\phi^n$. Due to the nonlinearity of the resulting equations, in our analysis we will primarily be interested in the asymptotic behavior of the model. For this, we will use methods of dynamical systems theory, as well as numerical integration and the representation of the model dynamics as a phase portrait (direction field). A review of the application of dynamical systems theory methods to cosmology and modified gravity can be found in \cite{Bahamonde2018}. 

\section{Action and Equations}

Consider a theory of gravity with a scalar field $\phi$ non-minimally coupled to curvature, described by an action $S$ of the following form\footnote{It should be emphasized that the sign of $\eta$ in the action (\ref{eq:action}) is chosen opposite to that in Refs. \cite{Sushkov2009,Saridakis2010,Sushkov2012,Skugoreva2013,Matsumoto2015,Matsumoto2018}.}

\begin{equation}\label{eq:action}
	S = \frac{1}{2}\int d^4x\sqrt{-g}\left[(M_{\text{Pl}}^2 - \xi\phi^2)R - (g^{\mu\nu} - \eta G^{\mu\nu})\phi_{,\mu}\phi_{,\nu} - 2V(\phi)\right] + S_m.
\end{equation}

Here $g_{\mu\nu}$ is the pseudo-Riemannian metric with signature $(-,+,+,+)$, $g = \det g_{\mu\nu}$, $R$ is the scalar curvature (the Ricci scalar), $M_{\text{Pl}}$ is the reduced Planck mass (equal to $\sqrt{\frac{c^4}{8\pi G}}$, where $G$ is Newton's gravitational constant, $c$ is the speed of light. From now on, we will work in the system of units $\hbar=c=8\pi G = 1$), $G_{\mu\nu}$ is the Einstein tensor, $V(\phi)$ is the scalar field potential, $\eta$ is the derivative coupling coefficient with $G_{\mu\nu}$ of dimension $m^{-2}$, $\xi$ is the dimensionless coupling coefficient of the scalar field with the curvature $R$. $S_m$ is the action for ordinary matter.

This action is a special case of Horndeski theory with the following choice of functions $G_i$: $G_2 = X - V(\phi),\ G_3 = 0,\ G_4 = \frac{1}{2}(M_{\text{Pl}}^2 - \xi\phi^2)R,\ G_5 = -\frac{1}{2}\eta\phi$.

By varying the action functional \eqref{eq:action} with respect to the metric $g_{\mu\nu}$ and the field $\phi$, we obtain the following system of differential equations, which are generalizations of the Einstein and Klein--Gordon equations, respectively: 

\begin{subequations}\label{eq:field}
	\begin{equation}\label{eq:g}
		G_{\mu\nu} = 
		T^{(m)}_{\mu\nu} + T^{(\phi)}_{\mu\nu} +
		\eta\Theta^{(1)}_{\mu\nu} + \xi\Theta^{(2)}_{\mu\nu},
	\end{equation}
	\begin{equation}\label{eq:phi}
		\square\phi - \eta G_{\mu\nu}\phi^{;\mu\nu} - \xi R\phi - V_\phi = 0, 
	\end{equation}
\end{subequations}

\noindent where $T^{(m)}_{\mu\nu}$ is the energy-momentum tensor of ordinary matter, $V_\phi \equiv dV(\phi)/d\phi$ and

\begin{eqnarray} 
	\label{eq:T^phi} 
	T^{(\phi)}_{\mu\nu} &=& 
	\phi_{,\mu}\phi_{,\nu} - 
	\frac{1}{2} \left(g_{\mu\nu} (\nabla \phi)^2 + V(\phi) \right), \\ \nonumber 
	\Theta^{(1)}_{\mu\nu} &=& 
	\frac{1}{2} R \phi_{,\mu}\phi_{,\nu} - 2\phi^{,\alpha} R_{\alpha(\mu}\phi_{,\nu)} - \phi^{,\alpha}\phi^{,\beta} R_{\alpha\mu\beta\nu} + \frac{1}{2} G_{\mu\nu} (\nabla \phi)^2 \\ \label{eq:Theta^1} 
	&&+\phi_{;\mu\nu}\square\phi - \phi^{;\alpha}_\mu\phi_{;\alpha\nu} - g_{\mu\nu} \left( \frac{1}{2} (\square\phi)^2 - \frac{1}{2} \phi^{;\alpha\beta}\phi_{;\alpha\beta} - R_{\alpha\beta}\phi^{,\alpha}\phi^{,\beta} \right), 
	\\ 
	\label{eq:Theta^2} 
	\Theta^{(2)}_{\mu\nu} &=& G_{\mu\nu}\phi^2 + g_{\mu\nu}\square(\phi^2) - (\phi^2)_{\mu\nu}.
\end{eqnarray}

By virtue of the Bianchi identities $\nabla^\mu G_{\mu\nu} = 0$ and the conservation law $\nabla^\mu T^{(m)}_{\mu\nu} = 0$ it follows from the equation \eqref{eq:g}

\begin{equation}\label{eq:g_seq} 
	\nabla^\mu \left( T^{(\phi)}_{\mu\nu} + \eta\Theta^{(1)}_{\mu\nu} + \xi\Theta^{(2)}_{\mu\nu} \right) = 0.
\end{equation}

Substituting here the expressions for $T^{(\phi)}_{\mu\nu}$, $\Theta^{(1)}_{\mu\nu}$ and $\Theta^{(2)}_{\mu\nu}$ from \eqref{eq:T^phi}, \eqref{eq:Theta^1}, and \eqref{eq:Theta^2}, we can verify that equations \eqref{eq:g} and \eqref{eq:phi} are not independent (they are equations with constraints).

Next, we will consider the resulting field equations in a homogeneous isotropic spatially flat metric:

\begin{equation}\label{eq:metric}
	ds^2 = -dt^2 + a^2(t)\left(dr^2 + r^2 d\Omega^2\right),
\end{equation}

\noindent where $a(t)$ is the scale factor, $d\Omega^2 = d\theta^2 + \sin^2\theta d\phi^2$ is the metric on the unit 2-sphere. Due to the homogeneity and isotropy of the model, it is also necessary to set $\phi = \phi(t)$ and $T^{(m)}_{\mu\nu} = \diag(\rho_m, P_m, P_m, P_m)$, where $\rho_m = \rho_m(t)$ and $P_m = P_m(t)$ are the energy density and the pressure of matter.

In the \eqref{eq:metric} metric, the $tt$ and $rr$ components of the Einstein equation \eqref{eq:g} have the following form:

\begin{subequations}\label{eq:metric_FE}
	\begin{equation}\label{eq:metric_FE_00}
		3H^2 = \rho_m + \rho_\phi ,
	\end{equation}
	\begin{equation}\label{eq:metric_FE_11}
		-(2\dot{H} + 3H^2) = P_m + P_\phi ,
	\end{equation}
\end{subequations}

\noindent where

\begin{subequations}
\begin{eqnarray}
	\rho_\phi & = & 
	\frac{1}{2}\dot{\phi}^2 + V(\phi) + \frac{9}{2}\eta H^2\dot{\phi}^2 
	+ 3\xi\{H^2\phi^2 + 2H\phi\dot{\phi}\},
	\label{eq:metric_rho}
	\\ 
	P_\phi & = & 
	\frac{1}{2}\dot{\phi}^2 - V(\phi) 
	-\frac{1}{2}\eta\left\{ ( 2\dot{H} + 3H^2)\dot{\phi}^2 + 4H\dot{\phi}\ddot{\phi} \right\}
	\nonumber\\ 
	& & -\xi\left\{ (2\dot{H} + 3H^2)\phi^2 + 4H\phi\dot{\phi} + 2\phi\ddot{\phi} + 2\dot{\phi}^2 \right\} 
\label{eq:metric_p} 
\end{eqnarray}
\end{subequations}
are effective energy density and scalar field pressure.

And the Klein--Gordon equation \eqref{eq:phi} is written as:

\begin{equation}\label{eq:metric_FE_phi}
	\ddot{\phi} + 3H\dot{\phi} + 3\eta\left\{ H^2\ddot{\phi} + 2H\dot{H}\dot{\phi} + 3H^3\dot{\phi} \right\}
	+ \xi\phi\{ \dot{H} + 2H^2 \} = -V_\phi
\end{equation}

Note that equations \eqref{eq:metric_FE_11} and \eqref{eq:metric_FE_phi} are second-order equations (i.e., dynamic equations), while \eqref{eq:metric_FE_00} is a first-order equation (i.e., first integral), the so-called Hamiltonian constraint.

We also write out the equation obtained by contracting \eqref{eq:g} with the inverse metric $g^{\mu\nu}$ (the trace of \eqref{eq:g}):

\begin{eqnarray}
	-R & = & -\rho_m + 3P_m - (\nabla \phi)^2 - 4V(\phi)
	\nonumber\\
	& & + \eta\left\{ -R_{\alpha\beta}\phi^{,\alpha}\phi^{,\beta} - \phi^{;\alpha\beta}\phi_{;\alpha\beta} + (\square\phi)^2\right\}
	+ \xi\left\{ -R\phi^2 + 3\square(\phi^2)\right\}.
\label{eq:g_tr}
\end{eqnarray}

In what follows, we will also analyze the behavior of the effective Planck mass, which in Horndeski theory can be defined as follows \cite{Bellini2014}:

\[
M_*^2(\phi,X,H) = 2(G_4 - 2XG_{4X} + XG_{5\phi} - \dot{\phi}HXG_{5X}).
\]

In our case, this expression takes the form

\begin{equation}
	\label{eq:effective_Planck_mass}
	M_*^2(\phi,\dot{\phi}) = M^2_{\text{Pl}}(1 - \xi\phi^2 - \eta\dot{\phi}^2/2).
\end{equation}

\section{Dynamical System}

\subsection{Equations and Notation}

To analyze the asymptotic behavior of the system, we rewrite the Friedmann equation using dimensionless variables. To do this, we divide the Friedman equation

\begin{equation}
	\label{eq:friedman2}
	3H^2F = \frac{1}{2}\dot{\phi}^2 + \frac{9}{2}\eta H^2\dot{\phi}^2 + 6\xi H\phi\dot{\phi} + V(\phi)
\end{equation}

\noindent by $3H^2F$, where for brevity we set $F = 1 - \xi\phi^2$, we get:

\begin{equation}\label{eq:dimless_constr}
	x + y + g + z = 1,
\end{equation}

\noindent where

\begin{equation}
	x = \frac{\dot{\phi}^2}{6H^2F}, \quad
	y = \frac{3\eta\dot{\phi}^2}{2F}, \quad
	g = \frac{2\xi\phi\dot{\phi}}{HF}, \quad
	z = \frac{V(\phi)}{3H^2F}.
\end{equation}

It is clear that $x$ and $z$ characterize the (normalized) magnitude of the kinetic and potential energy of the scalar field $\Omega_\phi = x + z$, while $y$ and $g$ characterize the derivative coupling $\Omega_\eta = y$ and the coupling with the scalar curvature $\Omega_\xi = g$, respectively.

To study the dynamics of these parameters, we differentiate them with respect to $N = \ln a$ (where $dN = d\ln a = Hdt$), obtaining:

\begin{subequations}\label{eq:dimless_sys1}
	\begin{align}
		x' &= x(2\delta + 2\varepsilon + g), \\
		y' &= y(2\delta + g), \\
		g' &= g(v + \delta + \varepsilon + g), \\
		z' &= z(\beta v + 2\varepsilon + g),
	\end{align}
\end{subequations}

\noindent where ${}' \equiv \frac{d}{dN}$. Here we introduced new parameters:

\begin{equation}
	v = \frac{\dot{\phi}}{H\phi}, \quad
	\delta = \frac{\ddot{\phi}}{H\dot{\phi}}, \quad
	\varepsilon = -\frac{\dot{H}}{H^2}, \quad
	\beta = \frac{\phi V_\phi}{V}
\end{equation}

\noindent and took advantage of the fact that $\frac{\dot{F}}{HF} = -g$.

Note that the expressions for the parameters $\delta$ and $\varepsilon$ coincide with the familiar slow-roll parameters. Moreover, the effective equation of state (barotropic index) $w_{eff} = -1 + \frac{2}{3}\varepsilon$.

We also note that the following relation holds:

\begin{equation}\label{eq:x_constr}
	g v = 12\xi x.
\end{equation}

To simplify further analysis, we set $V = V_0\phi^n$, then $\beta \equiv n$, where $n$ is a constant. Supplementing the system \eqref{eq:dimless_sys1} with an equation for $v$ and eliminating $x$ using \eqref{eq:x_constr} and $z$ using \eqref{eq:dimless_constr}, we obtain

\begin{subequations}\label{eq:dimless_sys2}
	\begin{align}
		y' &= y(2\delta + g), \\
		g' &= g(v + \delta + \varepsilon + g), \\
		\label{eq:dimless_sys2_v}
		v' &= v(\delta + \varepsilon - v),
	\end{align}
\end{subequations}

\noindent where $\delta$ and $\varepsilon$ are functions of $y$, $g$, and $v$ (the equations for $\delta$ and $\varepsilon$ will be obtained below).

However, the equations of the system \eqref{eq:dimless_sys2} are still not independent, since there is an algebraic constraint between the dimensionless variables \eqref{eq:dimless_constr}. Using the relation for the Hubble parameter, which we will need later,

\begin{equation}
	\label{eq:Hyx}
	H^2 = \frac{1}{9\eta}\cdot\frac{y}{x},
\end{equation}

\noindent and $\frac{\dot{\phi}^2}{H^2} = 6xF$, we obtain

\begin{equation}
	\label{eq:pre_constr}
	zv^ny = 3\eta V_0 \cdot 6^{\frac{n}{2}} \cdot x^{\frac{n}{2}+1} \cdot F^{\frac{n}{2}-1}.
\end{equation}

Thus, from the five equations \eqref{eq:dimless_sys1} and \eqref{eq:dimless_sys2_v}, using the three constraint equations \eqref{eq:friedman2}, \eqref{eq:x_constr}, and \eqref{eq:pre_constr}, only two independent equations remain, as expected.

Next, expressing the variable $x$ through the relation \eqref{eq:x_constr} and using the relation $F = 2v/(g + 2v)$ (from which we obtain that in the case $g = -2v$ the value of the field $\phi \rightarrow \infty$), we can finally write the relation \eqref{eq:pre_constr} as follows:

\begin{subequations}
	\label{eq:constr}
	\begin{align}
		zy(g + 2v)^{\frac{n}{2}-1} =
		A \cdot g^{\frac{n}{2}+1}, \quad & n > 2, \\
		zy = A \cdot g^2, \quad & n = 2, \\
		zy = A \cdot g^{\frac{n}{2}+1} \cdot (g + 2v)^{1-\frac{n}{2}}, \quad & -2 < n < 2, \\ 
		zy = A \cdot (g + 2v)^{2}, \quad & n = -2, \\ 
		zyg^{-1-\frac{n}{2}} = 
		A \cdot (g + 2v)^{1-\frac{n}{2}}, \quad & n < -2, 
	\end{align}
\end{subequations}

\noindent where $A = \frac{\eta V_0}{8\xi^{\frac{n}{2}+1}}$ --- constant. 

Different values of the parameter $n$ is used here to explicitly prevent zero from being raised to a negative power or part of the equation from vanishing.

To express $\delta$ and $\varepsilon$ in terms of $y$, $g$, and $v$, we divide the sum of the two Friedmann equations by $H^2F$ and multiply the Klein-Gordon equation by $\frac{\dot{\phi}}{H^3F}$, obtaining the following system of linear equations:

\begin{subequations}
	\begin{equation*}
		2\varepsilon = 6x + y(2 - \frac{4}{3}\delta + \frac{2}{3}\varepsilon) + g(1 - v - \delta),
	\end{equation*}
	\begin{equation*}
		6x\delta + 18x + y(6 + 2\delta - 4\varepsilon)
		+ g(6 - 3\varepsilon) + 3nvz = 0.
	\end{equation*}
\end{subequations}

Or, collecting the coefficients of $\delta$ and $\varepsilon$:

\begin{subequations}\label{eq:sys_XY}
	\begin{equation}
		(3g + 4y)\delta + (-2y + 6)\varepsilon = 3(6x + 2y + g - gv),
	\end{equation}
	\begin{equation}
		(-2y - 6x)\delta + (3g + 4y)\varepsilon = 3(6x + 2y + 2g + nvz),
	\end{equation}
\end{subequations}

\noindent where $z = 1 - y - x - g$ and $x = \frac{gv}{12\xi}$. From here, it is easy to obtain explicit expressions for $\delta$ and $\varepsilon$.

\subsection{Critical Points}

We will be primarily interested in the asymptotic behavior of the system near critical points. To find the critical points of the system, we must solve the equations \eqref{eq:dimless_sys2}, whose right-hand sides are set equal to zero. We will not eliminate the parameter $z$ from the equations using the constraint equation \eqref{eq:pre_constr}, but will proceed as follows. First, we will seek a general solution to the system of equations \eqref{eq:dimless_sys2}. Then, from \eqref{eq:constr}, we will determine the values of the parameter $n$ for which the obtained solutions satisfy the constraint \eqref{eq:pre_constr}. We substitute the obtained parameter values into the \eqref{eq:sys_XY} system and find the final values of the dimensionless parameters of the dynamic system (expressed in terms of the free parameters of the theory). Since the right-hand sides of the equations \eqref{eq:dimless_sys2} split into two parts, we need to analyze 8 cases depending on whether the variables $y$, $g$, and $v$ are zero:

Finally we get:

\begin{enumerate}
	\item $y=1,\ g=0,\ v=0,\ \delta=0,\ \varepsilon=\frac{3}{2},\ \forall n$; 
	\item $y=r_1,\ g=r_2,\ v=0,\ \delta=-\frac{r_2}{2},\ \varepsilon=-\frac{r_2}{2},\ \forall n$: 
	\begin{itemize}
		\item $r_1(1-r_1-r_2) = Ar_2^2,\ \text{where}\ r_1=-\frac{3}{2}\cdot\frac{r_2(r_2+4)}{r_2+6}$; 
	\end{itemize}
	\item $y=r_1,\ g=0,\ v=r_2,\ \delta=0,\ \varepsilon=r_2,\ n=-2$: 
	\begin{itemize}
		\item $4Ar_2^2 = r_1(1-r_1),\ \text{where}\ r_1 = \frac{3r_2}{r_2+3}$; 
	\end{itemize}
	\item $y=0,\ g=r,\ v=-\frac{r}{2},\ \delta=-\frac{r}{4},\ \varepsilon=-\frac{r}{4},\ n < 2$: 
	\begin{itemize} 
		\item $r = \frac{2(n-4)\xi}{(n+2)\xi-1}$; 
	\end{itemize}
	\item $y=0,\ g=r,\ v=-\frac{r}{2},\ \delta=\frac{r}{2}-3,\ \varepsilon=3-r,\ n < 2$: 
	\begin{itemize} 
		\item $r = 12\xi \pm 12\sqrt{\xi(\xi-\frac{1}{6})}$; 
	\end{itemize}
	\item $y=r_1,\ g=r_2,\ v=-\frac{r_2}{2},\ \delta=-\frac{r_2}{2},\ \varepsilon=0,\ n=2$: 
	\begin{itemize} 
		\item $r_1 = -\frac{3}{2}\cdot\frac{r_2^2(1-\frac{1}{4\xi})+r_2}{r_2+3}$, 
		$r_1(\frac{r_2^2}{24\xi}-r_2-r_1+1) = Ar_2^2$; 
	\end{itemize}
	\item $y=r_1,\ g=r_2,\ v=-\frac{r_2}{2},\ \delta=-\frac{r_2}{2},\ \varepsilon=0,\ n < 2$: 
	\begin{itemize} 
		\item $r_1 = 1 - r_2 + \frac{r_2^2}{24\xi}$, where 
		$\frac{r_2^3}{12\xi} + (1-\frac{1}{2\xi})r_2^2 - r_2 + 6 \equiv 0$. 
	\end{itemize}
\end{enumerate}

We excluded from consideration the cases $y=0,\ g=0,\ v=0,\ \delta=r,\ \varepsilon=0,\ \forall n$ and $y=0,\ g=0,\ v=r,\ \delta=r,\ \varepsilon=0,\ n=0$, which will be considered further in the \ref{sec:numeric} section, since they correspond to $\phi = \text{const}$ and require separate consideration. Cases 1 and 3 correspond to the dominance of the non-minimal derivative coupling ($g=0$), cases 4 and 5 --- non-minimal coupling with scalar curvature ($y=0$). For cases 2, 6, and 7, both non-minimal terms have comparable contributions to the dynamics. In this case, a non-degenerate ($y \neq 0$) de Sitter regime ($\varepsilon=0$) is possible only for cases 6 and 7.

Repeating the above steps for the case $V(\phi) \equiv 0$ (now $y = 1 - x - g$), we obtain:

\begin{enumerate}
	\item $g=0,\ v=0,\ \delta=0,\ \varepsilon=\frac{3}{2}$;
	\item $g=1,\ v=0,\ \delta=-3,\ \varepsilon=2$;
	\item $g=0,\ v=\frac{3}{2},\ \delta=0,\ \varepsilon=\frac{3}{2}$;
	\item $g=r_1,\ v=-\frac{r_1}{2},\ \delta=r_2,\ \varepsilon=-r_2-\frac{r_1}{2}$, where 
	\begin{equation} 
		\label{eq:eq4} 
		\begin{cases} 
			\frac{r_1^3}{24\xi} + \left(\frac{r_2}{4\xi} + \frac{1}{2\xi} - \frac{5}{2}\right)r_1^2 + (1 - 3r_2)r_1 - 6 = 0, \\ 
			-\frac{r_1^3}{12\xi} + \left(\frac{1}{2\xi} + \frac{1}{2}\right)r_1^2 + (3r_2-2)r_1 - 6r_2 - 6 = 0.
		\end{cases}
	\end{equation}
\end{enumerate}

\subsection{Stability and Asymptotics}

We will investigate the stability behavior of the dynamic system near the critical points found. To do this, we impose small perturbations to the values of the dynamic system parameters $y,\ g,\ v$ at the critical points. Then, in the linear approximation for small perturbations, the equations take the form (the equations for unperturbed values are satisfied identically):

\begin{equation}
	\frac{d}{dN}
	\left(
	\begin{array}{c}
		\delta y \\
		\delta g \\
		\delta v
	\end{array}
	\right)
	=
	M \times
	\left(
	\begin{array}{c}
		\delta y \\
		\delta g \\
		\delta v
	\end{array}
	\right)
\end{equation}

\noindent where $M$ is the Jacobian matrix $\frac{D(y',g',v')}{D(y,g,v)}$.

The stability of the dynamics near a critical point is determined by the eigenvalues $\lambda$ of the Jacobian matrix $M$ (which are the roots of the characteristic equation).

The most common types of critical points are as follows. A point is called a stable (unstable) node if all eigenvalues are real and negative (positive); a stable (unstable) focus if the eigenvalues are complex and their real parts are negative (positive); and a saddle if the eigenvalues have opposite signs. If any eigenvalue takes the value zero, further investigation is required to establish stability (for example, a critical point may be stable for one region of phase space and unstable for another).

At stationary points, the parameters of the dynamic system take constant values. Integrating the expression for $\varepsilon = -\frac{\dot{H}}{H^2}$, we obtain for the Hubble parameter:

\begin{subequations}
	\label{eq:H_asympt}
	\begin{equation}
		\label{eq:H_asympt_pow}
		H = \frac{1}{\varepsilon t}, \quad a \propto t^{1/\varepsilon}, \qquad \varepsilon \neq 0,
	\end{equation}
	\begin{equation}
		\label{eq:H_asympt_exp}
		H = \text{const}, \quad a \propto e^{Ht}, \qquad \varepsilon = 0,
	\end{equation}
\end{subequations}

\noindent where $t_0$ is the constant of integration (the beginning (the reference period of cosmological time) is set equal to zero.

It is clear that the asymptotic behavior of the Hubble parameter in the case $\varepsilon \neq 0$ is a power law. The universe expands with acceleration if $\varepsilon < 1$. Furthermore, in the case $\varepsilon < 0$, the Hubble parameter (scale factor) reaches infinite values in a finite time (the "Big Rip" singularity).

If $y \neq 0$, we can obtain an expression for the field $\phi(t)$ independently of the Hubble parameter, since $y = \frac{3\eta \dot{\phi}^2}{2(1-\xi\phi^2)}$ is independent of $H$. From this we obtain the differential equation

\[
\frac{\dot{\phi}^2}{1-\xi\phi^2} = \frac{2y}{3\eta}.
\]

Integrating which, we get:

\begin{enumerate}
	\item $y\eta > 0$: 
	\begin{subequations} 
		\label{eq:phi_y_1} 
		\begin{equation} 
			\phi = \pm\frac{1}{\sqrt{\xi}}\sin{\sqrt{\frac{2y\xi}{3\eta}}t}, \qquad \xi > 0, 
		\end{equation} 
		\begin{equation} 
			\phi = \pm\frac{1}{\sqrt{|\xi|}}\sinh{\sqrt{\frac{2y|\xi|}{3\eta}}t}, \qquad \xi < 0, 
		\end{equation} 
	\end{subequations}
	\item $y\eta < 0$:
	\begin{equation}
		\label{eq:phi_y_2}
		\phi = \pm\frac{1}{\sqrt{\xi}}\cosh{\sqrt{\frac{2y\xi}{3\eta}}t}, \qquad \xi > 0,
	\end{equation}
\end{enumerate}

\noindent where the integration constant $t_0$ is set equal to zero.

If $y=0$, but $g \neq 0$, then in the case of a power-law behavior \eqref{eq:H_asympt_pow} of the Hubble parameter, we have the differential equation

\[ \frac{2\xi\phi\dot{\phi}}{1-\xi\phi^2} = \frac{g}{\varepsilon t}, \]

\noindent integrating which, we obtain

\begin{equation}
	\label{eq:phi_g_pow}
	\phi = \pm \sqrt{\frac{1}{\xi}(1 - (Ct)^{-g/\varepsilon})},
\end{equation}

\noindent where $C$ is an arbitrary constant.

From this, we can also obtain the equation of the asymptotic trajectory in coordinates $(\phi,\dot{\phi})$. Calculating the derivative $\dot{\phi}$ and eliminating $t$ using the formula \eqref{eq:phi_g_pow}, we obtain

\begin{equation}
	\label{eq:traj_g_pow}
	\dot{\phi} = \frac{Cg}{2\xi\varepsilon}\cdot\frac{(1-\xi\phi^2)^{1+\frac{\varepsilon}{g}}}{\phi},
\end{equation}

\noindent where $C$ is a parameter numbering the curves in the family.

If the Hubble parameter is constant ($\varepsilon=0$), we have the differential equation

\[ \frac{2\xi\phi\dot{\phi}}{1-\xi\phi^2} = gH, \]

\noindent where, according to the formula \eqref{eq:Hyx}, $H^2 = \frac{1}{9\eta}\cdot\frac{1-g-x}{x}$ and $x=\frac{gv}{12\xi}$. Integrating it, we obtain

\begin{equation}
	\label{eq:phi_g_exp}
	\phi = \pm \sqrt{\frac{1}{\xi}(1 - Ce^{-gHt})},
\end{equation}

\noindent where $C$ is an arbitrary constant.

We obtain the equation for the asymptotic trajectory similarly to the case of the power-law behavior of the Hubble parameter \eqref{eq:traj_g_pow}:

\begin{equation}
	\label{eq:traj_g_exp}
	\dot{\phi} = \frac{gH}{2\xi}\cdot\frac{1-\xi\phi^2}{\phi}.
\end{equation}

\subsubsection{Case $z=0$}

Critical points and their properties in the absence of a scalar field potential ($V(\phi) \equiv 0$):

\paragraph{1.} 

$(g_*, v_*) = (0, 0)$. In this case, we obtain $y=1$. Eigenvalues: $[\frac{3}{2},\frac{3}{2}]$. The dynamics are dominated by the non-minimal derivative term $\Omega_\eta$. The critical point is an unstable node; in other words, this asymptotic behavior occurs when moving into the past. The effective equation of state $w_{eff} = 0$: the universe is effectively filled with matter (``dust'').

By the formula \eqref{eq:H_asympt} from $\varepsilon=\frac{3}{2}$, we obtain for the Hubble parameter and the scale factor
\[
H = \frac{2}{3t}, \quad a \propto t^{\frac{2}{3}}.
\]

Substituting the expressions for $\phi$ from \eqref{eq:phi_y_1} and \eqref{eq:phi_y_2}, we can verify that this stationary point exists ($v = \frac{\dot{\phi}}{H\phi}$ and $\delta = \frac{\ddot{\phi}}{H\dot{\phi}}$ tend to zero) only if $\eta > 0$ ($y = 1$) and $\phi_0 \neq 0$.

\paragraph{2.} 

$(g_*, v_*) = (1, 0)$. In this case, we obtain $y=0$. Eigenvalues: $[-5,-1]$. The non-minimal coupling term with curvature $\Omega_\xi$ dominates in the dynamics. The critical point is a stable node, so this asymptotic behavior is reached moving into the future. The effective equation of state is $w_{eff} = \frac{1}{3}$: the universe is effectively filled with radiation.

By formula \eqref{eq:H_asympt} from $\varepsilon=2$, we obtain for the Hubble parameter and scale factor
\[
H = \frac{1}{2t}, \quad a \propto t^{\frac{1}{2}}
\]

By formula \eqref{eq:phi_g_pow} from $g=1$, we obtain
\[
\phi = \pm \sqrt{\frac{1}{\xi} - \frac{1}{\xi\sqrt{Ct}}}.
\]

Since $t \rightarrow +\infty$, we see that this asymptotics exists only for $\xi > 0$. It is also clear that $\xi \rightarrow \pm\frac{1}{\sqrt{\xi}}$, from which, using \eqref{eq:effective_Planck_mass}, we obtain $M_* \rightarrow 0$ for the effective Planck mass.

From the formula \eqref{eq:traj_g_pow}, we obtain the equation of the trajectory in the $(\phi,\dot{\phi})$ plane near the stationary point:
\[ \dot{\phi} = \frac{C}{4\xi}\cdot\frac{(1-\xi\phi^2)^3}{\phi}. \]

\paragraph{3.} 

$(g_*, v_*) = (0, \frac{3}{2})$. In this case, we obtain $y=1$. Eigenvalues: $[-\frac{3}{2},3]$. The dynamics are dominated by the non-minimal derivative term $\Omega_\eta$. The critical point is a saddle point; in other words, the system avoids this point. The effective equation of state $w_{eff} = 0$: The universe is effectively filled with matter (``dust'').

Using the formula \eqref{eq:H_asympt} from $\varepsilon=\frac{3}{2}$, we obtain for the Hubble parameter and the scale factor
\[
H = \frac{2}{3t}, \quad a \propto t^{\frac{2}{3}}
\]

Substituting the expressions for $\phi$ from \eqref{eq:phi_y_1} and \eqref{eq:phi_y_2}, we can verify that this stationary point exists ($v = \frac{\dot{\phi}}{H\phi} = \frac{3}{2}$ and $\delta = \frac{\ddot{\phi}}{H\dot{\phi}} = 0$) only if $\eta > 0$ ($y=1$) and $\phi_0 = 0$.

\paragraph{4.}

$(g_*,v_*) = (r_1, -\frac{r_1}{2})$. In this case, we obtain $y = 1 - r_1 + \frac{r_1^2}{24\xi}$. Expressing $r_2$ from the second equation of the system \eqref{eq:eq4} and substituting it into the first, we obtain:

\begin{equation}
	\label{eq:eq4_expr}
	\begin{cases}
		r_2 = \frac{1}{6}(r_1-6)\left(\frac{r_1^2}{6\xi}-r_1-2\right)/(r_1-2), \\
		\left(\frac{r_1^2}{24\xi}-r_1+1\right)\left(\frac{r_1^3}{12\xi}+\left(1-\frac{1}{2\xi}\right)r_1^2-r_1+6\right)/(r_1-2)=0.
	\end{cases}
\end{equation}

The case $r_1=2$ is satisfied only when $\xi=\frac{1}{6}$, while the value of $r_2$ is undefined.

The second equation \eqref{eq:eq4_expr} is equivalent to the combination of two equations:

\begin{subequations}
	\begin{equation}
		\label{eq:eq4_quad}
		\frac{r_1^2}{24\xi}-r_1+1 = 0,
	\end{equation}
	\begin{equation}
		\label{eq:eq4_cub}
		\frac{r_1^3}{12\xi}+\left(1-\frac{1}{2\xi}\right)r_1^2-r_1+6 = 0.
	\end{equation}
\end{subequations}

The first of these equations corresponds to $y=0$ (the asymptotic value of the Hubble parameter is zero). Expressing $r_1^2$ from it and substituting it into the expression for $r_2$, we obtain $r_2 = \frac{r_1}{2}-3$. From this, $\varepsilon = -r_2-\frac{r_1}{2} = 3 - r_1$, where $r_1 = 12\xi \pm 12\sqrt{\xi(\xi-\frac{1}{6})}$ and $\xi \notin (0,\frac{1}{6})$. For $\xi \rightarrow \infty$, the asymptotic behavior of $r_*(\xi)$ is: $r_1 = 24\xi - 1$ and $r_1 = 1$.

\begin{figure}[htb]
	\centering
	\includegraphics[width=0.5\textwidth]{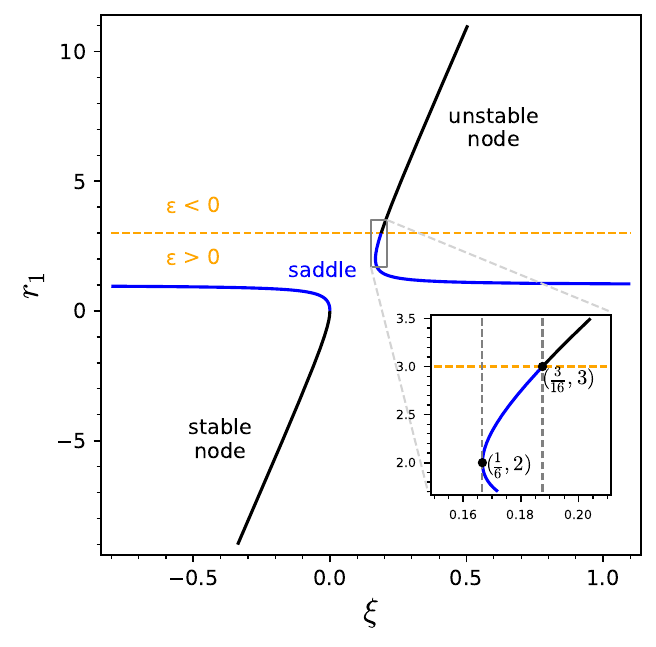}
	\caption{The $y=0$ curve (equation \eqref{eq:eq4_quad}). The blue part of the curve corresponds to saddle points, the black part to nodal points. The lower right corner shows a zoomed-in region near $\xi=\frac{1}{6},\ \frac{3}{16}$.}
	\label{fig:eq4quad}
\end{figure}

By numerically determining the eigenvalues for various parts of the $r_1(\xi)$ curve, we can verify that

\begin{itemize}
	\item for $r_1 = 12\xi - 12\sqrt{\xi(\xi-\frac{1}{6})}$, where $\xi < 0$, both eigenvalues are negative; thus, this part of the curve corresponds to a stable (attractive) node;
	\item for
	\begin{equation}
		\begin{cases}
			r_1 = 12\xi + 12\sqrt{\xi(\xi-\frac{1}{6})},\quad \xi \in (-\infty, 0) \cup (\frac{1}{6}, \frac{3}{16}), \\
			r_1 = 12\xi - 12\sqrt{\xi(\xi-\frac{1}{6})},\quad \xi > \frac{1}{6}.
		\end{cases}
	\end{equation}
	eigenvalues of different signs; thus, this part of the curve corresponds to a saddle point;
	\item for $r_1 = 12\xi + 12\sqrt{\xi(\xi-\frac{1}{6})}$, where $\xi > \frac{3}{16}$, both eigenvalues are positive; thus, this part of the curve corresponds to an unstable (repulsive) node.
\end{itemize}

The conclusions obtained above can also be verified analytically using the asymptotic behavior of the curve \eqref{eq:eq4_quad}. It can be shown that for $|\xi|\ \gg 1$, for the asymptotics $r_1 = 1$, the eigenvalues tend to $[-4,1]$, and for the asymptotics $r_1 = 24\xi-1$, they tend to $[24\xi-1, 48(\xi-\frac{1}{6})]$. Thus, for $\xi < 0$, the asymptotic expression $r_1 = 24\xi-1$ corresponds to a stable node, while for $\xi > 0$, it corresponds to an unstable node.

Expressing $r_1^3$ from the second equation similarly and substituting it into the expression for $r_2$, we obtain $r_2 = -\frac{r_1}{2}$. Hence, $\varepsilon = 0$. Thus, the solutions of the second equation yield the de Sitter asymptotic expression:

\[ H = \text{const},\quad a \propto e^{Ht}. \]

Calculating the discriminant of the cubic equation \eqref{eq:eq4_cub}, we obtain

\begin{equation}
	\label{eq:eq4_cub_discrim}
	D = \frac{3}{\xi^3} - \frac{20}{\xi^2} + \frac{79}{3\xi} - 23.
\end{equation}

The number of real roots of a cubic equation is determined by its discriminant as follows: for $D < 0$, there is one real root; for $D = 0$, there are two coinciding real roots and one distinct root (the case of three coinciding roots does not occur for this equation, as will be easily verified below); finally, for $D > 0$, there are three distinct real roots.

It is easy to show that the number of distinct real roots of the equation \eqref{eq:eq4_cub} is determined by the discriminant \eqref{eq:eq4_cub_discrim} as follows:

\begin{itemize}
	\item $D < 0 \iff \xi \in (-\infty,0) \cup (\xi_*,+\infty)$ --- 1 root;
	\item $D = 0 \iff \xi = \xi_*$ --- 2 roots;
	\item $D > 0 \iff \xi \in (0, \xi_*)$ --- 3 roots;
\end{itemize}

\noindent where

\begin{equation}
	\label{eq:xi_crit}
	\xi_* \approx 0.19 \iff \frac{1}{\xi_*} \approx 5.28.
\end{equation}

However, it should be kept in mind that if $y = 1 - r_1 + \frac{r_1^2}{24\xi} = 0$, then from the formula \eqref{eq:Hyx}, it follows $H=0$ (in this case $x = \frac{gv}{12\xi} = -\frac{r_1^2}{24\xi} \neq 0$), which means that the exponential asymptotics $a \propto e^{Ht}$ loses its meaning. Twice substituting $r_1^2 = 24\xi(r_1-1)$ into the cubic equation \eqref{eq:eq4_cub}, we obtain $r_1 = \frac{6-24\xi}{5-24\xi}$, substituting which into the previous relation, we obtain a quadratic equation on $\xi$. Solving it, we find the values of $\xi$ for which among the roots \eqref{eq:eq4_cub} there is one that sets $y$ (and therefore $H$) to zero:

\begin{equation}
	\label{eq:xi_H0}
	\xi = \frac{1}{6}, \quad \xi = \frac{3}{16}.
\end{equation}

Thus, for the given values of the parameter $\xi$, there are not three (since $\frac{1}{6}, \frac{3}{16} < \xi_*$) different de Sitter regimes, but only two.

Next, in order for the de Sitter asymptotics to make sense, we need to make sure that at the roots of the equation \eqref{eq:eq4_cub} the value of $H^2$ according to the formula \eqref{eq:Hyx} is positive, i.e. It is necessary to verify that the inequality is satisfied.

\[ H^2 = \frac{1}{9\eta}\cdot\frac{y}{x} > 0. \]

Expressing $y$ and $x$ here in terms of $g,\ v$ and substituting $g=r_1,\ v=-\frac{r_1}{2}$, we obtain:

\begin{equation}
	\label{eq:xi_ineq}
	\eta(r_1^2 - 24\xi \cdot r_1 + 24\xi) < 0 \iff \xi\eta y < 0,
\end{equation}

\noindent the left-hand side of which vanishes at $r_1 = 12\xi \pm 12\sqrt{\xi(\xi-\frac{1}{6})}$.

The cubic equation \eqref{eq:eq4_cub} is difficult to study analytically. Let's try to draw some conclusions based on a graphical representation of its solutions $r_*$ as functions of $\xi$.

\begin{figure}[htb]
	\centering
	\includegraphics[width=\textwidth]{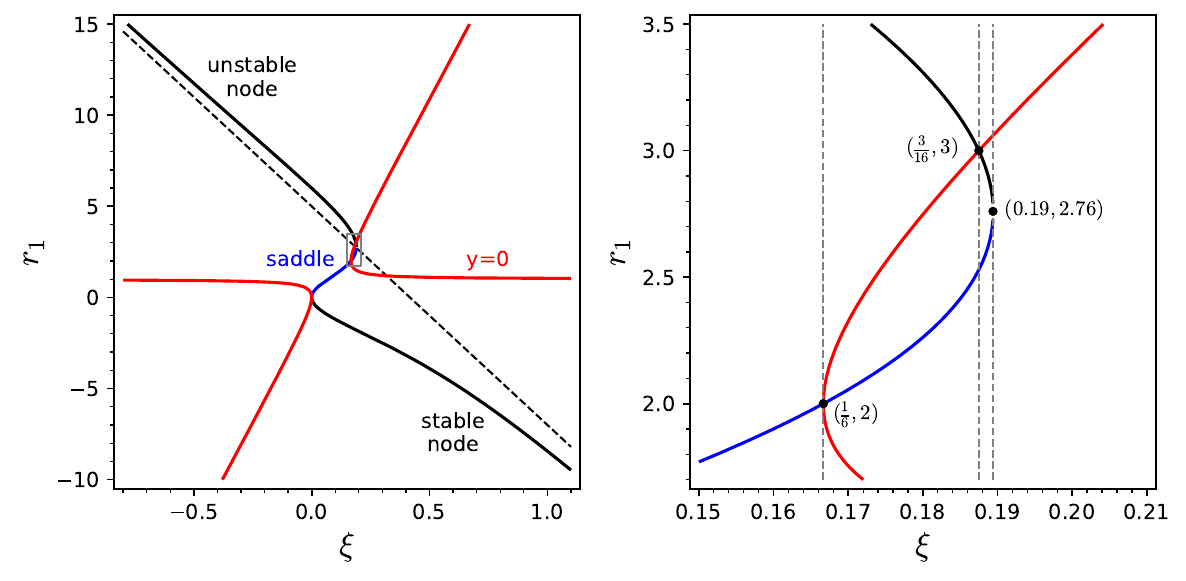}
	\caption{Left: The black and blue solid lines represent the roots $r_*$ of the equation \eqref{eq:eq4_cub} as functions of $\xi$; the black dashed line represents the asymptotic behavior of $r_* = -12\xi+5$ as $\xi \rightarrow \infty$; the red line represents the curve $y=0$ ($H=0$). For $\xi=0$, there is a single value $r_1=6$. Right: An enlarged region is shown for the values $\xi = \frac{1}{6},\ \frac{3}{16}$ and $\xi_* \approx 0.19$. The region inside the hyperbola $y=0$ for $\eta < 0$ corresponds to the asymptotics with $H^2 < 0$ and has no physical meaning.}
	\label{fig:eq4}
\end{figure}

Figure \ref{fig:eq4} shows that the distribution of the number of roots for different intervals of $\xi$ values was described correctly. It can be shown that the asymptotic behavior of the solution $r_1$ of the equation \eqref{eq:eq4_cub} for $\xi \rightarrow 0$ and $\xi \rightarrow \infty$ is given by:

\begin{equation}
	\label{eq:eq4_r1_asymp}
	\begin{cases}
		r_1(\xi) \approx -12\xi + 5,\quad \text{for}\ |\xi|\ \gg 1, \\
		r_1(\xi) \approx -12\xi + 6,\quad \text{for}\ |\xi|\ \ll 1, \\
		r_1(\xi) \approx \pm2\sqrt{3}\sqrt{\xi},\quad \text{for}\ |\xi|\ \ll 1\ (\xi \neq 0).
	\end{cases}
\end{equation}

In the same way, we can define the asymptotic value of $H^2$ for $\xi \rightarrow 0$ and $\xi \rightarrow \infty$:

\begin{equation}
	\label{eq:eq4_H2_asymp}
	\begin{cases}
		9\eta H^2 \approx -3 - \frac{1}{\xi},\quad \text{for}\ |\xi|\ \gg 1, \\
		9\eta H^2 \approx -1 + \frac{10}{3}\xi,\quad \text{for}\ |\xi|\ \ll 1, \\
		9\eta H^2 \approx -3 \pm \frac{10\sqrt{3}}{3}\sqrt{\xi},\quad \text{at}\ |\xi|\ \ll 1\ (\xi \neq 0). 
	\end{cases}
\end{equation}

Substituting $\xi$, expressed in terms of $r_1$, from equation~\eqref{eq:eq4_cub} into formula~\eqref{eq:Hyx} for the asymptotic value of $H^2$ at $y = 1 - r_1 + \frac{r_1^2}{24\xi}$ and $x = \frac{gv}{12\xi} = -\frac{r_1^2}{24\xi}$, we obtain an expression for the curve $9\eta H^2(\xi)$ in parametric form:

\begin{equation}
	\label{eq:eq4_H2_curve}
	\begin{cases}
		\xi = -\frac{r_1^2(r_1 - 6)}{12(r_1^2 - r_1 + 6)}, \\
		9\eta H^2 = -3 + \frac{12r_1}{r_1^2 - r_1 + 6}.
	\end{cases}
\end{equation}

\begin{figure}[htb]
	\centering
	\includegraphics[width=0.6\linewidth]{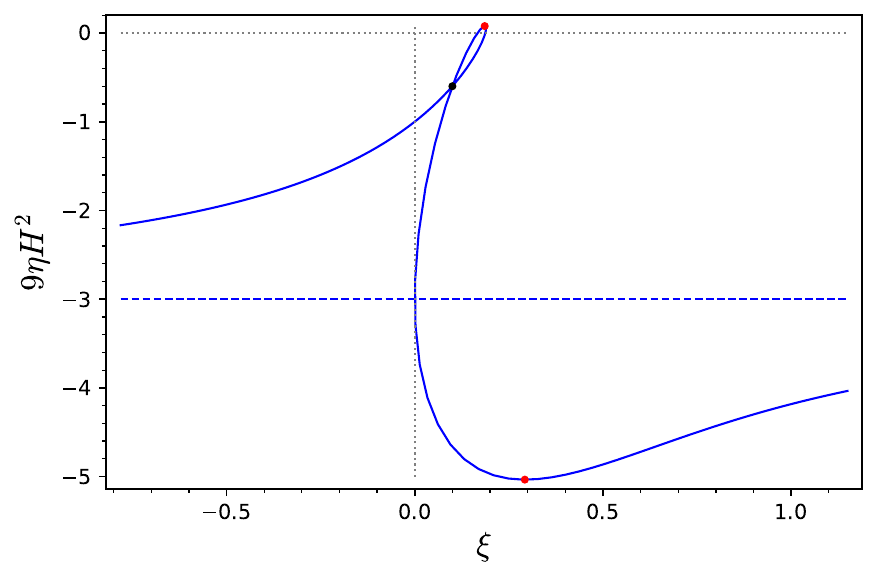}
	\caption{The value of $9\eta H^2$ depending on the value of the parameter $\xi$. The blue dashed line shows the asymptote $9\eta H^2 = -3$ at $\xi \rightarrow \infty$. The black dot shows the point of self-intersection of the curve with coordinates $(0.1, 0.6)$. The red dots mark the extremum points. The value $\xi=0$ corresponds to a single point on the curve with the value $9\eta H^2 = -1$.}
	\label{fig:eq4_H2}
\end{figure}

The minimum value of $9\eta H^2$ is reached at $\xi = \frac{6 + \sqrt{6}}{2(\sqrt{6} + 12)} \approx 0.292$ ($r_1 = -\sqrt{6}$) and is equal to $-\frac{36 + 15\sqrt{6}}{12 + \sqrt{6}} \approx -5.030$. The maximum of $9\eta H^2$ is achieved at $\xi = \frac{6 - \sqrt{6}}{2(\sqrt{6} - 12)} \approx 0.185$ ($r_1 = \sqrt{6}$) and is equal to $-\frac{36 + 15\sqrt{6}}{12 + \sqrt{6}} \approx 0.078$. In addition, the curve $9\eta H^2(\xi)$ has a self-intersection at $\xi = \frac{1}{10}$ ($r_1 = 3 \pm \sqrt{3}$), in which two of the three asymptotic values of $9\eta H^2$ coincide with each other and are equal to $-\frac{6}{10}$. It is obvious that the value of $H^2$ for $\xi \rightarrow \infty$ tends to $\frac{1}{3|\eta|}$, and for $\xi \rightarrow 0$ --- to $\frac{1}{9|\eta|}$, which is consistent with the results obtained in the works ~\cite{Sushkov2009, Saridakis2010}.

Comparing the solutions $r_1(\xi)$ with the inequality \ref{eq:xi_ineq}, we find that the exponential asymptotics $H=\text{const}$ exists for $\eta > 0$ only over a narrow interval (see \ref{eq:xi_crit} and \ref{eq:xi_H0}):

\[ \frac{1}{6} < \xi < \xi_* \approx \frac{1}{5.28}.\]

By numerically determining the eigenvalues for three different branches of the curve $r_1(\xi)$, we can verify that

\begin{itemize}
	\item for the branch $\xi \in (-\infty,\xi_*)$, both eigenvalues are positive, thus this branch corresponds to an unstable (repulsive) node;
	\item for the branch $\xi \in (0,\xi_*)$, the eigenvalues have opposite signs, thus this branch corresponds to a saddle point;
	\item for the branch $\xi \in (0,+\infty)$, both eigenvalues are negative, thus this branch corresponds to a stable (attracting) node.
\end{itemize}

The conclusions obtained above can also be verified analytically using the asymptotic behavior of the curve \eqref{eq:eq4_cub}. It can be shown that for $|\xi|\ \gg 1$, for the asymptotics $r_1 = r_* = -12\xi+5$, the eigenvalues tend to the values $[-12\xi+2,\ -12\xi+5]$. Thus, indeed, for $\xi \rightarrow -\infty$ the asymptotic behavior corresponds to an unstable node, and for $\xi \rightarrow \infty$ -- to a stable one.

From this it can be seen that in the case $\eta > 0$ there is a very narrow interval of $\xi$ values for which the exponential asymptotics, in addition to the saddle point, also has a point of the repulsive node type, namely (see \eqref{eq:xi_crit} and \eqref{eq:xi_H0}):

\[ \frac{3}{16} < \xi < \xi_* \approx \frac{1}{5.28}.\]

Thus, the values $0,\ \frac{1}{6},\ \frac{3}{16},\ \frac{1}{5.28}$ of the parameter $\xi$ divide the possible phase portraits of the system in the case of stationary points with $z = 0$ and $H = \text{const}$ into different classes (see figure \ref{fig:eq4}), which will be considered in section \ref{sec:numeric}. In this case, for $\eta < 0$, the following classes of phase portraits actually exist based on the type and number of attractors (see Fig. \ref{fig:plt_arr}):

\begin{figure}[htb]
	\centering
	\includegraphics[width=\linewidth]{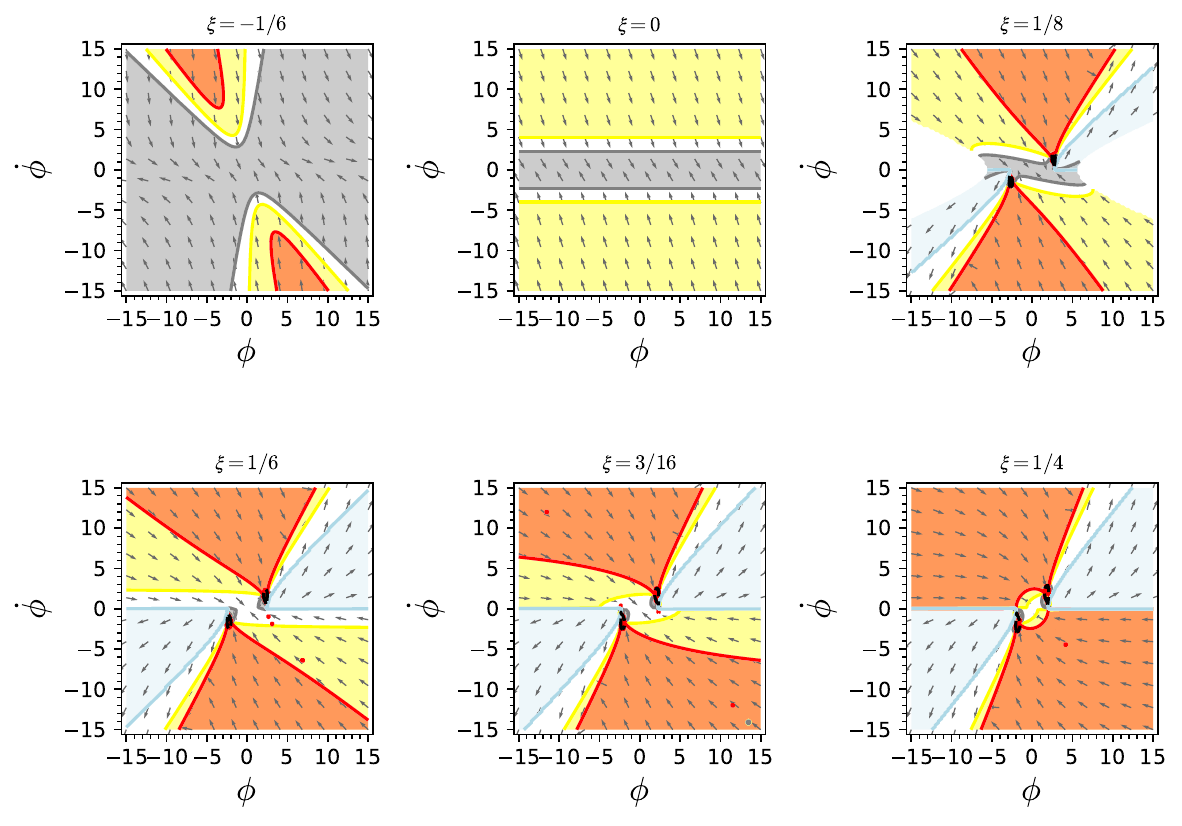}
	\caption{Different classes of phase portraits for $V(\phi) \equiv 0$. Phase portraits of the system in $(\phi, \dot{\phi})$ coordinates for $\eta < 0$. White and gray areas correspond to expansion with deceleration (barotropic index $w \leq 1/3$ and $w > 1/3$, respectively), yellow and red areas correspond to accelerated ($w < -1/3$) and superaccelerated expansion ($w < -1,\ \dot{H} > 0$). The black dotted lines indicate the discontinuity of the dynamics (the $\ddot{\phi}$ singularity). The light blue regions correspond to the contraction $H < 0$. The black dots indicate stationary points $\pm 1/\sqrt{\xi}$. The white regions without arrows indicate unphysical regions.}
	\label{fig:plt_arr}
\end{figure}

\begin{itemize}
	\item $\xi < 0$: $w_{\text{eff}} = -1$ (unstable node), $w_{\text{eff}} > 1$ (stable node and saddle);
	\item $\xi = 0$: $w_{\text{eff}} = -1$ (unstable node), $w_{\text{eff}} = 1$ (stable node);
	\item $0 < \xi < \frac{1}{6}$: $w_{\text{eff}} = -1$ (unstable node, stable node, saddle), $w_{\text{eff}} = \frac{1}{3}$ (stable node);
	\item $\frac{1}{6} < \xi < \frac{3}{16}$: $w_{\text{eff}} = -1$ (unstable node, stable node), $w_{\text{eff}} < -1$ (unstable node), $w_{\text{eff}} = \frac{1}{3}$ (stable node);
	\item $\xi > \frac{3}{16}$: $w_{\text{eff}} < -1$ (unstable node), $w_{\text{eff}} = -1$ (stable node), $w_{\text{eff}} = \frac{1}{3}$ (stable node).
\end{itemize}

The type of stationary point for $\xi = \frac{1}{6}$ and $\xi = \frac{3}{16}$ is complex, since they belong to both branches of the solutions.

\subsubsection{Case $z \neq 0$}

The case $z \ne 0$ is much more difficult to analyze exhaustively than the case $z = 0$, including stability analysis (due to the presence of many parameters $\xi$, $\eta$, $V_0$, and $n$). Therefore, here we will primarily analyze the number of stationary regimes, indicating the behavior of the scale factor (parameter $\varepsilon$). Points 6 and 7 will be discussed in somewhat more detail due to the greater interest in these regimes.

\paragraph{1.}

$(y_*, g_*, v_*) = (1,0,0)$. Eigenvalues: $[\frac{3}{2},3]$. The dynamics are dominated by the non-minimal derivative term $\Omega_\eta$. The critical point is an unstable node; in other words, this asymptotic behavior occurs when moving into the past. The equation of state $w_{eff} = 0$: the universe is effectively filled with matter (``dust'').

Using the formula \eqref{eq:H_asympt} from $\varepsilon=\frac{3}{2}$, we obtain for the Hubble parameter and the scale factor

\[
H = \frac{2}{3t}, \quad a \propto t^{\frac{2}{3}}.
\]

Substituting the expressions for $\phi$ from \eqref{eq:phi_y_1} and \eqref{eq:phi_y_2}, we can verify that this stationary point exists ($v = \frac{\dot{\phi}}{H\phi}$ and $\delta = \frac{\ddot{\phi}}{H\dot{\phi}}$ tend to zero) only if $y\eta > 0$ and $\phi_0 \neq 0$.

\paragraph{2.}

$y=r_1,\ g=r_2,\ v=0,\ \delta=-\frac{r_2}{2},\ \varepsilon=-\frac{r_2}{2},\ \forall n$, where 

\begin{equation} 
	\begin{cases} 
		r_1=-\frac{3}{2}\cdot\frac{r_2(r_2+4)}{r_2+6},\\ 
		r_1(1-r_1-r_2) = Ar_2^2. 
	\end{cases}
\end{equation}

\noindent where $A = \frac{\eta V_0}{8\xi^{\frac{n}{2}+1}}$. Substituting the first into the second, we obtain

\[  r_2[(4A+3)r_2^3 + 6(8A+3)r_2^2+12(12A+5)r_2+144]/(r_2+6)^2 = 0. \]

As $A \rightarrow \infty$, the expression in square brackets tends to $4Ar_2(r_2+6)^2$. Thus, in the limit $A \rightarrow \infty$, the solutions tend to the values $r_*=0$ and $r_*=-6$. $A \rightarrow \infty$ only for $\xi \rightarrow 0,\ n > -2$. Since we exclude the limit $\xi=0$ from consideration, for all finite $A$, $r_2 \ne -6$. Furthermore, we will exclude the solution $r_2=0$, since in this case all parameters of the dynamic system except $z$ are equal to zero ($\phi = \text{const}$). Therefore, we need to consider the following equation:

\begin{equation}
	\label{eq:eqz2}
	(4A+3)r_2^3 + 6(8A+3)r_2^2+12(12A+5)r_2+144 = 0.
\end{equation}

Equation \eqref{eq:eqz2} is cubic and has an analytical solution $r_*(A)$. However, an explicit solution is cumbersome, so we will rely on a graphical representation of the solutions of equation \eqref{eq:eqz2}.

First, let's determine the number of roots depending on the value of $A$. Calculating the discriminant, we obtain:

\begin{equation}
	\label{eq:eqz2_discrim}
	D = -1728(48A + 25)(144A^2 - 136A + 33).
\end{equation}

Its only (real) root is $A=-\frac{25}{48}$. Thus, the number of roots of equation \eqref{eq:eqz2} is delimited by a single value of $A$. However, when $A = -\frac{3}{4}$, the cubic equation \eqref{eq:eqz2} reduces to a quadratic equation with two roots: $r_* = \frac{2}{3}(-2 \pm \sqrt{22})$.

Thus, we finally have:

\begin{itemize}
	\item $A < -\frac{25}{48} \iff D > 0$ --- 3 roots;
	\item $A = -\frac{3}{4} \iff D > 0$ --- 2 roots;
	\item $-\frac{3}{4} < A < -\frac{25}{48} \iff D > 0$ --- 3 roots;
	\item $A = -\frac{25}{48} \iff D = 0$ --- 2 roots;
	\item $A > -\frac{25}{48} \iff D < 0$ --- 1 root;
\end{itemize}

\begin{figure}[htb] 
	\centering 
	\includegraphics[width=\textwidth]{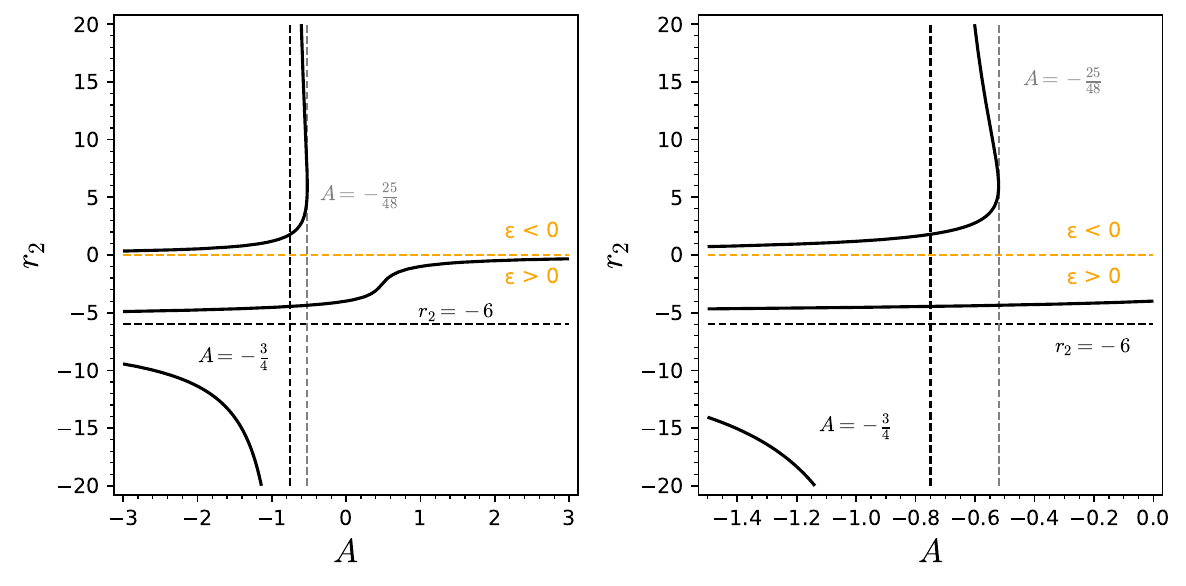} 
	\caption{Equation \ref{eq:eqz2}.} 
	\label{fig:eqz2}
\end{figure}

\paragraph{3.}

$y=r_1,\ g=0,\ v=r_2,\ \delta=0,\ \varepsilon=r_2,\ n=-2$, where

\begin{equation} 
	\begin{cases} 
		r_1 = \frac{3r_2}{r_2+3}, \\ 
		r_1(1-r_1) = 4Ar_2^2, 
	\end{cases}
\end{equation}

\noindent and $A = \frac{\eta V_0}{8\xi^{\frac{n}{2}+1}}$. Substituting the first equation into the second, we obtain

\[ r_2[4Ar_2^3 + 24Ar_2^2 + 6(6A+1)r_2 - 9]/(r_2+3)^2 = 0. \]

As $A \rightarrow \infty$, the expression in square brackets tends to $4Ar_2(r_2+3)^2$. Thus, in the limit $A \rightarrow \infty$, the solutions $r_2$ tend to the values $r_*=0$ and $r_*=-3$. Moreover, for all finite $A$, the value $r_2\ne -3$. Furthermore, we exclude the solution $r_2=0$, since in this case all parameters of the dynamical system except $z$ are equal to zero ($\phi = \text{const}$). Thus, we need to consider the following equation:

\begin{equation}
	\label{eq:eqz3}
	4Ar_2^3 + 24Ar_2^2 + 6(6A+1)r_2 - 9 = 0.
\end{equation}

Equation \eqref{eq:eqz3} is cubic and has an analytical solution $r_*(A)$. However, an explicit solution is cumbersome, so we will rely on a graphical representation of the solutions of equation \eqref{eq:eqz3}.

First, let's determine the number of roots depending on the value of $A$. Calculating the discriminant, we obtain:

\begin{equation}
	\label{eq:eqz3_discrim}
	D = -432A(9A + 8)(48A+1).
\end{equation}

The equation \eqref{eq:eqz3_discrim} has three real roots: $A=0,\ A=-\frac{8}{9},\ A = -\frac{1}{48}$. Thus, the number of roots of the equation \eqref{eq:eqz3} is limited by three values of $A$. Note that when $A=0$, the cubic equation \eqref{eq:eqz3} reduces to a linear equation with a single root $r_*=\frac{3}{2}$.

Taking this into account, we finally have:

\begin{itemize}
	\item $A < -\frac{8}{9} \iff D > 0$ --- 3 roots;
	\item $A = -\frac{8}{9} \iff D = 0$ --- 2 roots;
	\item $-\frac{8}{9} < A < -\frac{1}{48} \iff D < 0$ --- 1 root;
	\item $A = -\frac{1}{48} \iff D = 0$ --- 2 roots;
	\item $-\frac{1}{48} < A < 0 \iff D > 0$ --- 3 roots;
	\item $A \geq 0 \iff D < 0$ --- 1 root;
\end{itemize}

\begin{figure}[htb] 
	\centering 
	\includegraphics[width=\textwidth]{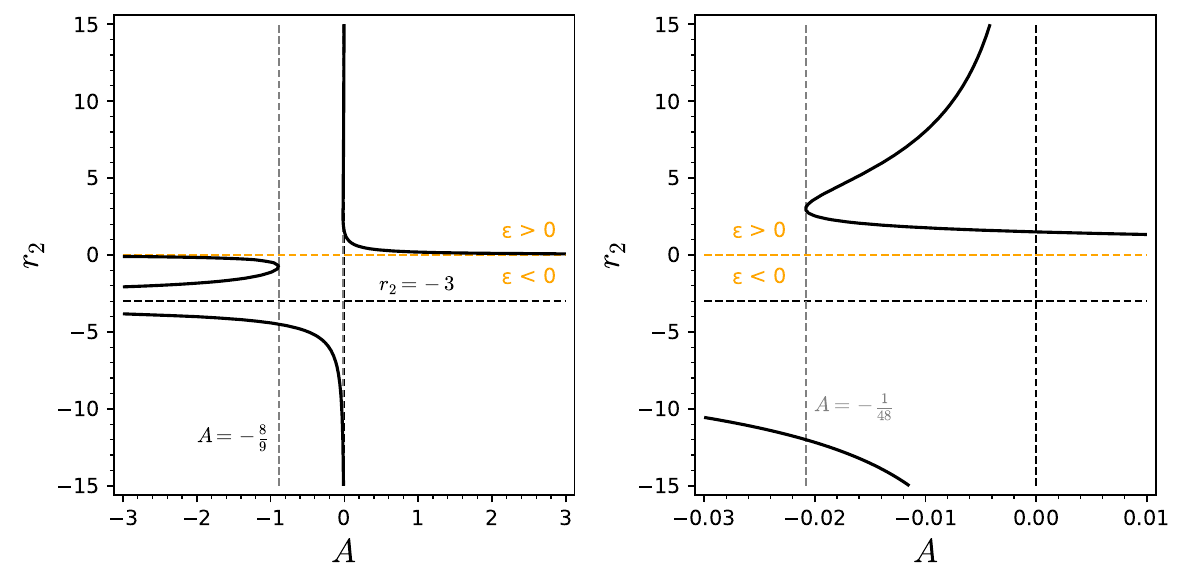} 
	\caption{Equation \ref{eq:eqz3}.} 
	\label{fig:eqz3}
\end{figure}

\paragraph{4.}

$y=0,\ g=r,\ v=-\frac{r}{2},\ \delta=-\frac{r}{4},\ \varepsilon=-\frac{r}{4},\ n < 2$, where
\begin{equation} 
	\label{eq:eqz4} 
	r = \frac{2(n-4)\xi}{(n+2)\xi-1}.
\end{equation}

At $|\xi|\ \gg 1$ and $n \ne -2$, the value of $r$ quickly tends to a constant value $r = \frac{2(n-4)}{n+2}$. For $n=-2$, the hyperbola \eqref{eq:eqz4} degenerates into a straight line $r = -12\xi$.

\paragraph{5.}

$y=0,\ g=r,\ v=-\frac{r}{2},\ \delta=\frac{r}{2}-3,\ \varepsilon=3-r,\ n < 2$, where
\begin{equation}
	\label{eq:eqz5}
	r = 12\xi \pm 12\sqrt{\xi\left(\xi-\frac{1}{6}\right)}.
\end{equation}

A dependence of the form \eqref{eq:eqz5} has already been analyzed in the case 4 $z=0$ (see Fig.~\ref{fig:eq4quad}).

\paragraph{6.}

$y=r_1,\ g=r_2,\ v=-\frac{r_2}{2},\ \delta=-\frac{r_2}{2},\ \varepsilon=0,\ n=2$, where
\begin{equation*}
	\begin{cases}
		r_1 = -\frac{3}{2}\cdot\frac{r_2^2(1-\frac{1}{4\xi})+r_2}{r_2+3}, \\
		r_1(\frac{r_2^2}{24\xi}-r_2-r_1+1) = Ar_2^2.
	\end{cases}
\end{equation*}

Substituting the first equation into the second and $\frac{\eta V_0}{8\xi^2}$ in place of $A$, we obtain the following equation for $r_2$:

\begin{equation}
	\label{eq:eq6}
	-\frac
	{((4\xi-1)r_2 + 4\xi)(r_2^3 + 6(2\xi-1)r_2^2 -12\xi r_2 + 72\xi)}
	{8r_2(r_2+3)^2} = \eta V_0.
\end{equation}

\noindent where $V_0 = \frac{m^2}{2}$.

In general, this is a fourth-degree equation. Its exact solution is cumbersome, and for specific parameter values, it is better to solve it numerically. However, we can solve it approximately if we assume that $|\eta V_0|\ \ll 1$. In the limit $\eta V_0 \rightarrow 0$, we obtain the root $-\frac{4\xi}{4\xi-1}$ and the cubic equation \eqref{eq:eq4_cub}, which we have already considered. Its root for $|\xi|\ \gg 1$ has the dependence $r_2 \approx -12\xi+5$, and for $|\xi| \ll 1$, the root of the upper (unstable branch) is equal to $r_2 \approx 6(1-2\xi)$. Under the given assumptions, we can write an approximate solution in the form $r_2 = r_2^{(0)} + \alpha$, where $r_2^{(0)}$ is a solution to the ``homogeneous'' ($\eta V_0 = 0$) equation, and $\alpha$ is a small parameter. Substituting this expression into the original equation and expanding it in $\alpha$ up to the first nonzero term (a term of degree zero is identically zero by definition), we obtain

\begin{equation*}
	\begin{cases}
		r_2 \approx -\frac{4\xi}{4\xi-1} + \frac{2(8\xi-3)\eta m^2}{(4\xi-1)(6\xi-1)(16\xi-3)}, \\
		r_2 \approx 5 - 12\xi - \frac{\eta m^2}{\xi},\quad |\xi|\ \gg 1, \\
		r_2 \approx 6 - 12\xi + 9\eta m^2,\quad |\xi|\ \ll 1.
	\end{cases}
\end{equation*}

In the case of $\xi \in {\frac{1}{4}, \frac{1}{6}, \frac{3}{16}}$ the first approximate formula loses its meaning, and for $\xi = \frac{3}{8}$, equation \eqref{eq:eq6} degenerates into the equation

\[
-\frac{2r_2^2 - 9r_2 + 18}{32r_2} = \eta V_0,
\]

\noindent which has real roots only for $\eta V_0 \in (-\infty,\frac{3}{32}) \cup (\frac{21}{32}, \infty)$. Therefore, the specified values of $\xi$ require separate consideration.

Substituting the obtained expressions into the formula \eqref{eq:Hyx}, which at the stationary point under consideration takes the form

\[
H^2 = \frac{1}{\eta}\cdot\frac{(4\xi-1)r_2 + 4\xi}{r_2(r_2+3)},
\]

\noindent we obtain

\begin{equation}
	\label{eq:eq6_H2}
	\begin{cases}
		H^2 \approx -\frac{(4\xi-1)^2m^2}{2\xi(6\xi-1)(16\xi-3)}, \\
		H^2 \approx -\frac{1}{3\eta}\left( 1 + \frac{1}{3\xi} \right),\quad |\xi|\ \gg 1, \\
		H^2 \approx -\frac{1}{9\eta}\left( 1 - \frac{10}{3}\xi \right) + \frac{m^2}{9},\quad |\xi|\ \ll 1.
	\end{cases}
\end{equation}

It is clear that, to a first approximation, the first asymptotic behavior of $H^2$ does not depend on the non-minimal derivative coupling parameter $\eta$, and the Hubble parameter has a physical meaning ($H^2 \ge 0$) only for $\xi \in (-\infty,0) \cup (\frac{1}{6}, \frac{3}{16})$. It can be seen that the second asymptotic value of $H^2$ in the strong coupling regime $|\xi|\ \gg 1$ differs from the value of $H^2$ in the weak coupling regime $|\xi|\ \ll 1$ by approximately three times and, to a first approximation, does not depend on $V_0$.

\paragraph{7.}

$y=r_1,\ g=r_2,\ v=-\frac{r_2}{2},\ \delta=-\frac{r_2}{2},\ \varepsilon=0,\ n < 2$, where
\begin{equation*}
	\begin{cases}
		r_1 = 1 - r_2 + \frac{r_2^2}{24\xi}, \\
		\frac{r_2^3}{12\xi} + (1-\frac{1}{2\xi})r_2^2 - r_2 + 6 \equiv 0.
	\end{cases}
\end{equation*}

It was already shown above that the asymptotic value of $r_2$ for $|\xi|\ \ll 1$ is $r_2 \approx -12\xi + 6$, and for $|\xi|\ \gg 1$: $r_2 \approx -12\xi+5$. Substituting these expressions into the formula \eqref{eq:Hyx}, which at the stationary point under consideration takes the form

\[
H^2 = -\frac{r_2^2 - 24\xi r_2 + 24\xi}{r_2^2},
\]

\noindent we obtain

\begin{equation}
	\label{eq:eq7_H2}
	\begin{cases}
		H^2 \approx -\frac{1}{3\eta}\left( 1 + \frac{1}{3\xi} \right),\quad |\xi|\ \gg 1, \\
		H^2 \approx -\frac{1}{9\eta}\left( 1 - \frac{10}{3}\xi \right),\quad |\xi|\ \ll 1.
	\end{cases}
\end{equation}

It is evident that the obtained asymptotic values of the Hubble parameter coincide with those discussed above for the case $n=2$ in the limit $V_0 \rightarrow 0$. We again note that the value of $H^2$ in the strong-coupling regime $|\xi|\ \gg 1$ is approximately three times larger than the value of $H^2$ in the weak-coupling regime $|\xi|\ \ll 1$.

\section{Numerical Analysis}\label{sec:numeric}

Let us solve the system of equations \eqref{eq:metric_FE},\eqref{eq:metric_FE_phi} for $\ddot{\phi}$ and $\dot{H}$. We get:

\begin{subequations}\label{eq:dyn_sys} 
	\begin{equation}\label{eq:phi_tt} 
		\ddot{\phi} = \frac{P(H, \phi, \dot{\phi})}{Q(H, \phi, \dot{\phi})}, 
	\end{equation} 
	\begin{equation}\label{eq:H_t} 
		\dot{H} = \frac{R(H, \phi, \dot{\phi})}{Q(H, \phi, \dot{\phi})}, 
	\end{equation}
\end{subequations}

\noindent where

\begin{subequations} 
	\begin{eqnarray} 
		P(H, \phi, \dot{\phi}) &=& 
		\eta H \left( 1 - 6\xi + \frac{3}{2}(1 - 3\eta H^2) \right)\dot{\phi}^3 
		- \left(-\frac{\eta}{2}V_\phi - \xi\phi\left(1 - 6\xi - 21\eta H^2\right)\right)\dot{\phi}^2 
		\nonumber\\
		&&
		- 3H\left(1 - \eta\left\{H^2(1 - \xi\phi^2) - \frac{V(\phi)}{3}\right\} - \left[-\eta V(\phi) + \xi(1-6\xi)\phi^2\right]\right)\dot{\phi} 
		\nonumber\\
		&&+ \xi\phi(\phi V_\phi - 4V(\phi)) - V_\phi , 
		\\ 
		R(H, \phi, \dot{\phi}) &=& 
		- \frac{1}{3}\left(1 + 3\eta H^2\right)\left(1 - 6\xi + 9\eta H^2\right)\dot{\phi}^2 
		- 2\eta H\left(V_\phi + 12\xi H^2\phi\right)\dot{\phi} 
		\nonumber\\
		&&
		+ 4\xi H^2\left(1 - 6\xi + 3\eta H^2\right)\phi^2 
		- 4\left(1 + 3\eta H^2\right)\left(H^2 - \frac{V(\phi)}{3}\right) - 2\xi\phi V_\phi , 
		\\
		Q(H, \phi, \dot{\phi}) &=& 
		1 - \xi(1-6\xi)\phi^2 + 3\eta H^2(1 - \xi\phi^2) 
		+ 12\eta\xi H\phi\dot{\phi} 
		- \frac{\eta}{2} \left(1 - 9\eta H^2\right)\dot{\phi}^2 . 
	\end{eqnarray}
\end{subequations}

However, we only need to express $\ddot{\phi}$, and as the second independent equation we can take \eqref{eq:metric_FE_00}, which is an algebraic (quadratic) equation with respect to $H$:

\begin{equation}\label{eq:H}
	H(\phi, \dot{\phi}) =
	\frac{\xi\phi\dot{\phi} \pm \sqrt{D(\phi, \dot{\phi})}}{1 - \xi\phi^2 - \frac{3}{2}\eta\dot{\phi}^2} ,
\end{equation}

\noindent where

\begin{equation*}
	D(\phi, \dot{\phi}) =
	(\xi\phi\dot{\phi})^2 + \frac{1}{3}\left(\frac{1}{2}\dot{\phi}^2 + V(\phi)\right)\left(1 - \xi\phi^2 - \frac{3}{2}\eta\dot{\phi}^2\right) .
\end{equation*}

At points where the denominator in \eqref{eq:H} is zero, from the Friedmann equation \eqref{eq:metric_FE_00} for the Hubble parameter we obtain:

\[ H(\phi, \dot{\phi}) = -\frac{\frac{1}{2}\dot{\phi}^2 + V(\phi)}{6\xi\phi\dot{\phi}} .\]

Next, substituting the expression for $H$ \eqref{eq:H} into the equation \eqref{eq:phi_tt}, we obtain a closed system of differential equations for $\phi$ and $\dot{\phi}$.

It should be noted that in reality, the system \eqref{eq:dyn_sys} is a two-dimensional surface in the three-dimensional space $(\phi, \dot{\phi}, H)$, and the system's dynamics are represented by a line on this surface. This should be kept in mind, since in what follows we will consider the direction field of the dynamical system \eqref{eq:dyn_sys} projected onto the plane $H = 0$.

Thus, the equation \eqref{eq:phi_tt} for the scalar field $\phi$, given the relation \eqref{eq:H}, represents an autonomous dynamical system in normal form, i.e. resolved with respect to the highest derivative.

The relations $g = 0 (\Rightarrow x = 0)$ and $y = 0$ imply $\dot{\phi} = 0$, i.e., $\phi = \text{const}$. Thus, from the first Friedman equation \eqref{eq:friedman2}, we obtain

\begin{equation}
	\label{eq:H_stat}
	3H^2 = \frac{V(\phi)}{1 - \xi\phi^2}.
\end{equation}

Using this relation, we obtain a condition on the values of the field $\phi$ at stationary points from the requirement $\dot{\phi} = \ddot{\phi} = 0$:

\[
\frac{\xi\phi(\phi V_\phi - 4V(\phi)) - V_\phi}{1 - \xi(1-6\xi)\phi^2 + 3\eta V(\phi)} = 0.
\]

Thus, for the case of a power-law potential $V = V_0\phi^n$, this equation takes the following form:

\[
\frac{\phi^{n-1}(\xi(n-4)\phi^2 - n)}{1 - \xi(1-6\xi)\phi^2 + 3\eta V_0\phi^n} = 0.
\]

Hence, for $\xi \neq 0$, we obtain the value of $\phi$ in stationary points (with a non-zero denominator):

\begin{equation*}
	\begin{cases}
		\phi = 0,\quad n > 1, \\
		\phi^2 = \frac{n}{(n-4)\xi},\quad \forall n, \\
		\phi \rightarrow \infty,\quad n < 1 \cup n = 4.
	\end{cases}
\end{equation*}

Here, the symbol $\infty$ acts as an abbreviation for $\pm\infty$, and for the existence of stationary points $\phi = \pm\sqrt{\frac{n}{(n-4)\xi}}$, the expression under the radical obviously must be non-negative.

Now we determine stability near the stationary points found. To do this, we represent the value of the field $\phi$ near these points in the form $\phi = \phi_* + \delta\phi$ and linearize the system, i.e. consider it in the linear approximation in $\delta\phi$:

\[
\frac{d}{dt}
\left(
\begin{array}{c}
	\delta\phi \\ \delta\dot{\phi}
\end{array}
\right) =
\left(
\begin{array}{cc}
	0 & 1 \\
	\frac{\partial \ddot{\phi}}{\partial \phi} & \frac{\partial \ddot{\phi}}{\partial \dot{\phi}}
\end{array}
\right)
\left(
\begin{array}{c}
	\delta\phi \\ \delta\dot{\phi}
\end{array}
\right),
\]

Thus, for stability in our case, we have the following obvious conditions:

\begin{equation}\label{eq:stat_cond}
	\frac{\partial \ddot{\phi}}{\partial \phi} < 0, \quad
	\frac{\partial \ddot{\phi}}{\partial \dot{\phi}} < 0.
\end{equation}

To calculate the values of the partial derivatives of $\ddot{\phi}$ at a stationary point, we use the fact that $\ddot{\phi}=0$ there, for example:

\[
\frac{\partial \ddot{\phi}}{\partial \phi} =
\frac{\partial}{\partial \phi}\left(\frac{P}{Q}\right) = \frac{P'_\phi}{Q} - \ddot{\phi} \cdot \frac{Q'_\phi}{Q} = \frac{P'_\phi}{Q}
\]

From here (using the relation \eqref{eq:H_stat}) we obtain the following conditions for the stability of motion near a stationary point:

\begin{subequations}\label{eq:stat_cond_explicit}
	\begin{align}
		\label{eq:d2phi_dphi}
		\frac{\partial \ddot{\phi}}{\partial \phi} &=
		\frac{\xi(\phi^2 V_{\phi\phi} - 2\phi V_\phi - 4V(\phi)) - V_{\phi\phi}}{1 - \xi(1-6\xi)\phi^2 - \eta V(\phi)} < 0,
		\\
		\label{eq:d2phi_d1phi}
		\frac{\partial \ddot{\phi}}{\partial \dot{\phi}} &=
		\mp \sqrt{\frac{3 V(\phi)}{1-\xi\phi^2}} = -3H < 0
	\end{align}
\end{subequations}

Note that the relation \eqref{eq:d2phi_d1phi} can be obtained directly from the Klein-Gordon equation \eqref{eq:metric_FE_phi}, where $\dot{H} = 0$ at the stationary point. And the condition \eqref{eq:d2phi_dphi} for the power-law potential can be rewritten as

\begin{equation}
	\label{eq:stability_power_law}
	\frac{\xi(n+1)(n-4)\phi^n - n(n-1)\phi^{n-2}}{1 - \xi(1-6\xi)\phi^2 - \eta V_0\phi^n}V_0 < 0,\qquad n \ne 0,
\end{equation}

Now we determine the stability of the stationary points $\phi=0$. For odd $n > 0$, this point (obviously) turns out to be unstable (the sign of $\frac{\partial \ddot{\phi}}{\partial \phi}$ changes depending on the sign of $\phi$). For even $n > 0$, the value of $\frac{\partial \ddot{\phi}}{\partial \phi}$ at the point $\phi=0$ vanishes, but in the neighborhood of this point it behaves as $\sim -C\phi^{n-2} < 0$, where $C$ is a positive constant, i.e., the point is stable\footnote{This could be shown more rigorously using the Lyapunov method, where the Hubble parameter $H$ is taken as the Lyapunov function, and its time derivative is determined by the expression~\eqref{eq:H_t}. It can be shown that in a sufficiently small neighborhood of zero ($\phi \sim \dot{\phi} \sim 0$), the derivative of the Lyapunov function $\dot{H} \le 0$, which confirms the stability of the stationary point.}. For $n = 0$, the \eqref{eq:stability_power_law} condition for $\phi = 0$ takes the following form:

\[ \frac{\xi\Lambda}{1 - \eta\Lambda} > 0,\quad n = 0, \]

\noindent where $V_0 \equiv \Lambda$ is the cosmological constant.

The stability condition \eqref{eq:stability_power_law} for the infinitely distant stationary point $n = 4$ (assuming $V_0 > 0$) and the physicality condition for the Hubble parameter ($H^2 \ge 0$) from ~\eqref{eq:H_stat} imply, respectively:

\[
\eta < 0,\ \xi < 0,\quad n = 4.
\]

For infinitely distant stationary points $\phi \rightarrow \infty$ for $n < 1$, we obtain:

\[
\begin{cases}
	\frac{(n+1)(n-4)}{1 - 6\xi}\phi^{n-2} > 0, \quad &n \ne -1, \\
	\frac{2}{\xi(1-6\xi)}\phi^{-5} < 0, \quad &n = 1.
\end{cases}
\]

\noindent which, for the point at infinity $\phi \rightarrow +\infty$ (for even $n$, the obtained conclusions will also remain valid for $\phi \rightarrow -\infty$, and for odd $n$, the conditions for the stability of $\phi \rightarrow -\infty$ will be the opposite) takes the following form:

\begin{equation}
	\begin{cases}
		\xi > \frac{1}{6},\quad &-1 < n < 1 \\
		0 < \xi < \frac{1}{6},\quad &n = -1 \\
		\xi < \frac{1}{6},\quad &n < -1.
	\end{cases}
\end{equation}

For stationary points $\phi^2 = \frac{n}{(n-4)\xi}$, stability analysis for arbitrary $n$ is difficult. However, for $n = 2$, the existence and stability condition takes the form:

\[ \frac{1 - \sqrt{1 + 6\eta V_0}}{6} < \xi < 0,\quad \eta > 0. \]

Thus, in the case $n = 2,\ \xi < 0,\ \eta < 0$, which we will consider in more detail below, the stationary points $\phi^2 = \frac{1}{-\xi}$ turn out to be unstable (saddle points).

\subsection{Case $\textbf{z=0}$}

Let us analyze cosmological dynamics using the example of a theory with conformal constraint $\xi = 1/6$. Solving the equation \eqref{eq:eq4_cub} for a given value of $\xi$, we obtain, according to \eqref{eq:xi_crit}, three roots:

\[ r_1=2,\ r_1=1 + \sqrt{7},\ r_1 = 1 - \sqrt{7}. \]

However, according to \eqref{eq:xi_H0}, for $r_1=2$, the asymptotics loses meaning (the asymptotic value of the Hubble parameter is zero), and this value must be excluded.

For $r_1 = 1 \pm \sqrt{7}$, we have the following eigenvalues.

\[ [0.65, 3.65],\ [-4.65, -1.65]. \]

Thus, the first eigenvalue $r_1 = 1 + \sqrt{7}$ corresponds to the repulsive node, and in the second eigenvalue $r_1 = 1 - \sqrt{7}$ is an attractive node.

From \eqref{eq:xi_ineq}, we see that the de Sitter asymptotics exists ($H^2 > 0$) only for $\eta < 0$. Therefore, for further analysis, we choose the value $\eta = -1/9$. Then, using the formula \eqref{eq:Hyx}, we obtain:

\[ H^2 = \frac{4 - \sqrt{7}}{4 + \sqrt{7}},\quad H^2 = \frac{4 + \sqrt{7}}{4 - \sqrt{7}}. \]

Whence, we have the following asymptotic values of the Hubble parameter:

\[ H \approx 0.45, \quad H \approx 2.22. \]

\begin{figure}[htb]
	\centering
	\begin{minipage}[t]{0.58\textwidth}
		\centering\includegraphics[width=\textwidth]{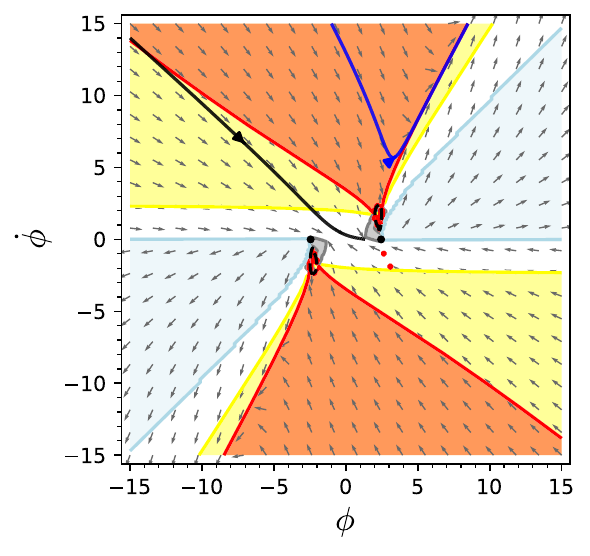}
	\end{minipage}
	\hfill
	\begin{minipage}[t]{0.41\textwidth}
		\centering\includegraphics[width=\textwidth]{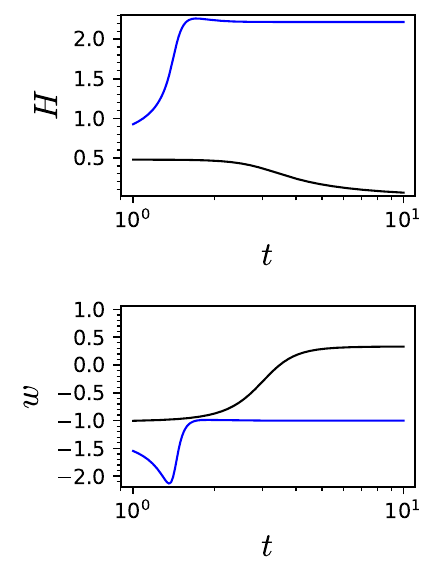}
	\end{minipage}
	\caption{Example of the model dynamics for $\eta=-1/9,\ \xi=1/6$ and initial values $\phi=-15, \dot{\phi}=14$ (black curve) and $\phi=-1, \dot{\phi}=15$ (blue curve). \emph{Left.} Phase portrait of the system in $(\phi, \dot{\phi})$ coordinates. White and gray areas correspond to expansion with deceleration (barotropic index $w \leq 1/3$ and $w > 1/3$, respectively), yellow and red areas correspond to accelerated ($w < -1/3$) and superaccelerated expansion ($w < -1,\ \dot{H} > 0$). Black dotted lines show the violation of the continuity of the dynamics (singularity $\ddot{\phi}$). Light blue areas correspond to contraction $H < 0$. Black dots denote stationary points $\pm 1/\sqrt{\xi}$. \emph{Right.} Dynamics of the Hubble parameter $H$ and the effective equation of state $w$. It is clear that after the inflationary stage, the universe transits to a radiation-dominated stage with $w=\frac{1}{3}$.}
\end{figure}

\subsection{The Case for Potential}

\begin{figure}[htb]
	\centering
	\begin{minipage}[t]{0.58\textwidth}
		\centering\includegraphics[width=\textwidth]{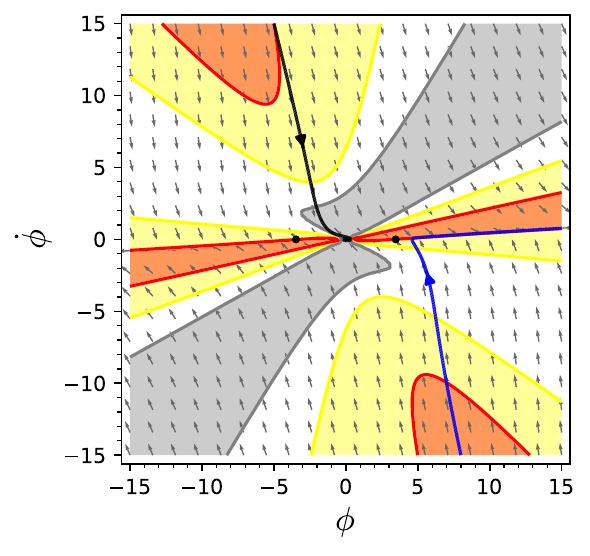}
	\end{minipage}
	\hfill
	\begin{minipage}[t]{0.41\textwidth}
		\centering\includegraphics[width=\textwidth]{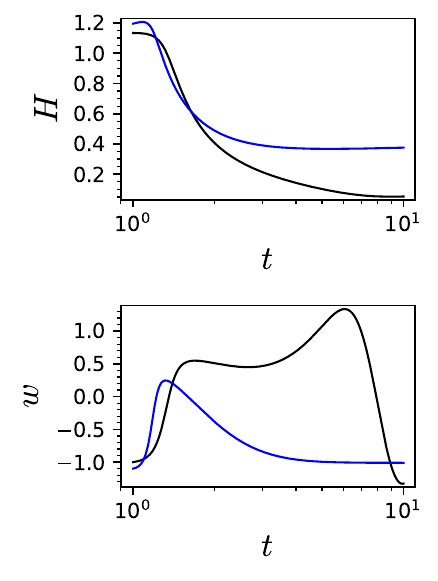}
	\end{minipage}
	\caption{Example of the model dynamics for $\eta=-1/9,\ \xi=-1/12,\ m=1/3$ and initial values $\phi=-5, \dot{\phi}=15$ (black curve) and $\phi=8, \dot{\phi}=-15$ (blue curve). \emph{Left.} Phase portrait of the system in $(\phi, \dot{\phi})$ coordinates. White and gray areas correspond to expansion with deceleration (barotropic index $w \leq 1/3$ and $w > 1/3$, respectively), yellow and red areas correspond to accelerated ($w < -1/3$) and superaccelerated expansion ($w < -1,\ \dot{H} > 0$). Black dots denote saddle points $\pm 1/\sqrt{|\xi|}$. \emph{Right.} Dynamics of the Hubble parameter $H$ and the effective equation of state $w$. It is clear that after the inflationary stage, the universe either transits to an oscillatory regime (black curve) or enters a new quasi-de Sitter stage (blue curve).}
	\label{fig:numeric_z}
\end{figure}

Let us analyze cosmological dynamics using the example of a theory with $\xi < 0,\ \eta < 0$ and a quadratic potential $V(\phi) = \frac{1}{2}m^2\phi^2$. We set $\eta = -\frac{1}{9}$, $\xi = -\frac{1}{12}$, and $m = \frac{1}{3}$. Then we have $|\xi|\ \ll 1$ and $\eta V_0 = \frac{1}{2}\eta m^2 = \frac{1}{2}\cdot\frac{1}{9^2} \ll 1$ and can apply the formulas for the approximate value of the Hubble parameter \eqref{eq:eq6_H2} from the previous section. We obtain the values

\[
H \approx 0.427,\quad H \approx 1.136,
\]

\noindent which are in good agreement with the numerical calculations (Fig.~\ref{fig:numeric_z}).

\section{Conclusion}

In this paper, we investigated the dynamics of homogeneous isotropic cosmological models with a spatially flat metric in gravity theories with a scalar field non-minimally coupled to the curvature. The non-minimal coupling here was characterized by the presence of terms of the form $\xi R\phi^2$ and $\eta G^{\mu\nu}\phi_{,\mu}\phi_{,\nu}$ (derivative coupling) in the action functional. We also considered a theory that includes a power-law scalar field potential. Due to the nonlinearity of the resulting dynamic equations, in our analysis we were primarily interested in the asymptotic behavior and also used numerical integration, including to represent the model dynamics as a phase portrait.

The presence of a non-minimal coupling of the form $\xi R\phi^2$ in a theory with a kinetic coupling $\eta G^{\mu\nu}\phi_{,\mu}\phi_{,\nu}$ qualitatively changes the nature of the dynamics compared to a theory characterized only by a derivative coupling \cite{Sushkov2009, Saridakis2010}. Specifically, for the case $V(\phi) \equiv 0$, an interesting stable regime with $w_{eff} = \frac{1}{3}$ was obtained, as well as stable and unstable (depending on the non-minimal coupling parameter $\xi$) stages of quasi-de Sitter and phantom expansions. Moreover, for the interval $0 < \xi < \frac{3}{16}$, both unstable stationary and stable quasi-de Sitter regimes ($\eta < 0$) exist. 

The presence of the potential $V(\phi)$ of the power-law scalar field $V_0\phi^n$ sharply increases the possible stationary regimes, but these are mainly phantom ($w_{eff} < -1$) and extremely stiff ($w_{eff} > 1$) expansions. For $n \le 2$, the influence of the scalar field potential on the unstable quasi-de Sitter stage asymptotically vanishes, and the dependence of the Hubble parameter $H(\xi)$ tends to that at $V(\phi) \equiv 0$. Moreover, for $n = 2$, there is a transition between the unstable and stable quasi-de Sitter stages, separated by an expansion with $w_{eff} > 0$. 

\section*{Acknowledgments}
The work is supported by the Russian Science Foundation grant 
No.~25-22-00163.


\end{document}